\documentstyle[epsfig]{cupconf}

\ifoldfss
\else
  \ifnfssone
    \newmathalphabet{\mathit}
      \addtoversion{normal}{\mathit}{cmr}{m}{it}
      \addtoversion{bold}{\mathit}{cmr}{bx}{it}
    \newmathalphabet{\mathcal}
      \addtoversion{normal}{\mathcal}{cmsy}{m}{n}
    \else
    \ifnfsstwo
    \fi
  \fi
\fi

\def\hexnumber#1{\ifcase#1 0\or1\or2\or3\or4\or5\or6\or7\or8\or9\or
 A\or B\or C\or D\or E\or F\fi }

\makeatletter
\ifx\CUP@mtlplain@loaded\undefined
\else
\fi
\makeatother
 \makeatletter
 \ifx\CUP@mtlplain@loaded\undefined
   \font\tenbmi=cmmib10 at 10pt
   \font\sevenbmi=cmmib10 at 7pt
   \font\fivebmi=cmmib10 at 5pt

   \newfam\bmifam
   \textfont\bmifam=\tenbmi
   \scriptfont\bmifam=\sevenbmi
   \scriptscriptfont\bmifam=\fivebmi
   
 \fi
 \makeatother
\ifnfsstwo

\fi
\ifnfssone

\fi
\ifoldfss

\fi

\mathchardef\varLambda="0103

\makeatletter
\ifx\CUP@mtlplain@loaded\undefined
\else
\fi
\makeatother
\makeatletter
\ifx\CUP@mtlplain@loaded\undefined
  \font\tenbms=cmbsy10
  \font\sevenbms=cmbsy10 at 7pt
  \font\fivebms=cmbsy10 at 5pt
  \newfam\bmsfam
  \textfont\bmsfam=\tenbms
  \scriptfont\bmsfam=\sevenbms
  \scriptscriptfont\bmsfam=\fivebms

  \edef\bsy@{\hexnumber\bmsfam}
  \mathchardef\bnabla="0\bsy@72
\fi
\makeatother
\title[Nucleosynthesis]{Nucleosynthesis Basics and Applications
to Supernovae}

\author[Friedrich-Karl Thielemann {\it et al.\/}]%
{F.\ls-\ls K.\ns T\ls H\ls I\ls E\ls L\ls E\ls M\ls A\ls N\ls 
N$^{1,5}$\ns\\
T.\ns R\ls A\ls U\ls S\ls C\ls H\ls E\ls R$^1$,\ns\\
C.\ns F\ls R\ls E\ls I\ls B\ls U\ls R\ls G\ls H\ls A\ls U\ls S$^{1,5}$,\ns
K.\ns N\ls O\ls M\ls O\ls T\ls O$^{2,5}$,\ns\\
M.\ns H\ls A\ls S\ls H\ls I\ls M\ls O\ls T\ls O$^3$,\ns\\
B.\ns P\ls F\ls E\ls I\ls F\ls F\ls E\ls R$^4$\ns
\and \ns 
K.\ls-\ls L.\ns K\ls R\ls A\ls T\ls Z$^4$}

\affiliation{$^1$Departement f\"ur Physik und Astronomie, Universit\"at Basel,
CH-4056 Basel, Switzerland\\[\affilskip]
$^2$Department of Astronomy and Research Center for the Early Universe,
University of Tokyo, Tokyo 113, Japan\\[\affilskip]
$^3$Department of Physics, Faculty of Science, Kyushu University,
Fukuoka 810, Japan\\[\affilskip]
$^4$Institut f\"ur Kernchemie, Universit\"at Mainz, D-55128 Mainz,
Germany\\[\affilskip]
$^5$Institute for Theoretical Physics, University of California,
Santa Barbara, CA 93106-4030}

\begin{document}
\ifnfssone
\else
  \ifnfsstwo
  \else
    \ifoldfss
      \let\mathcal\cal
      \let\mathrm\rm
      \let\mathsf\sf
    \fi
  \fi
\fi

\maketitle

\begin{abstract}
This review concentrates on nucleosynthesis processes in general and
their applications to massive stars and supernovae. A brief
initial introduction is given to the physics in astrophysical plasmas which
governs composition changes. We present the basic equations for thermonuclear
reaction rates and nuclear reaction networks. The required nuclear physics
input for reaction rates is discussed, i.e. cross sections for nuclear
reactions, photodisintegrations, electron and positron captures, neutrino
captures, inelastic neutrino scattering, and beta-decay half-lives. 
We examine especially the present state of uncertainties in predicting 
thermonuclear reaction rates, while the status of experiments is discussed
by others in this volume (see M. Wiescher). It follows 
a brief review of hydrostatic burning stages in stellar evolution before
discussing the fate of massive stars, i.e. the nucleosynthesis in type II 
supernova explosions (SNe II). Except for SNe Ia, which are explained by 
exploding white dwarfs in binary stellar systems (which will not be discussed 
here), all other supernova types seem to be linked
to the gravitational collapse of massive stars (M$>$8M$_\odot$) at the
end of their hydrostatic evolution. SN1987A, the first type II supernova
for which the progenitor star was known, is used as an example
for nucleosynthesis calculations. 
Finally, we discuss the production of heavy elements in the r-process
up to Th and U and its possible connection to supernovae.
\end{abstract}

\firstsection 
\section{Thermonuclear Rates and Reaction Networks}

In this section we want to outline the essential features 
of thermonuclear reaction rates and nuclear reaction
networks. This serves the purpose to define a unified terminology
to be used throughout the review, more detailed discussions can be found
in Fowler, Caughlan, \& Zimmerman (1967,1975), Clayton (1983), Rolfs
\& Rodney (1988), Thielemann, Nomoto, \& Hashimoto (1994), and Arnett (1996).

\subsection{Thermonuclear Reaction Rates}

The nuclear cross section for a reaction between target $j$ and projectile $k$
is defined by 
\begin{equation}
\sigma = {\rm{number\ of\ reactions\ target^{-1} sec^{-1}} \over
{flux\ of\ incoming\ projectiles}} = {{r/n_j} \over {n_k v}}. 
\end{equation}
The second equality holds for the case that the relative velocity between
targets with the number density $n_j$ and projectiles with number
density $n_k$ is constant and has the value $v$. Then $r$, the number of 
reactions per cm$^3$ and sec, can be expressed as $r=\sigma v n_j n_k$.
More generally, when targets and projectiles follow specific distributions,
$r$ is given by

\begin{equation}
r_{j,k}=\int \sigma \vert \vec v_j -\vec v_k\vert d^3 n_j d^3 n_k.
\end{equation}

The evaluation of this integral depends on the type of particles and
distributions which are involved. For nuclei $j$ and $k$ in an astrophysical 
plasma, obeying a Maxwell-Boltzmann distribution,

\begin{equation}
d^3n_j=n_j ({{m_j} \over {2\pi kT}})^{3/2} {\rm exp}(- {{m_jv_j^2}
\over {2kT}}) d^3v_j,
\end{equation}
Eq.(1.2) simplifies to $r_{j,k}=<\sigma v> n_j n_k$. The thermonuclear reaction 
rates have the form (Fowler, Caughlan, \& Zimmerman 1967, Clayton 1983)

\begin{subeqnarray}
r_{j,k} & = &<\sigma v>_{j,k} n_j n_k \\
<j,k>:& = &<\sigma v>_{j,k}=({8 \over {\mu \pi}})^{1/2} (kT)^{-3/2}
\int_0 ^\infty E \sigma (E) {\rm exp}(-E/kT) dE. 
\end{subeqnarray}

Here $\mu$ denotes the reduced mass of the target-projectile system.
In astrophysical plasmas with high densities and/or low temperatures,
effects of electron screening become highly important. 
This means that the reacting nuclei, due to the background of electrons and
nuclei, feel a different Coulomb repulsion than in the case of bare nuclei. 
Under most conditions (with non-vanishing temperatures)
the generalized reaction rate integral can be separated into the traditional
expression without screening [Eq.(1.4)] and a screening factor (see e.g.
Salpeter \& van Horn 1969, Itoh, Totsuji, \& Ichimaru 1977, Hansen, Torrie, 
\& Veillefosse 1977, Alastuey \& Jancovici 1978, Itoh et al. 1979, Ichimaru, 
Tanaka, Iyetomi 1984, Ichimaru \& Utsumi 1983, 1984, Thielemann \& Truran 1987,
Fushiki \& Lamb 1987, Itoh et al. 1990, Schramm \& Koonin 1990,
Ichimaru 1993, Chabrier \& Schatzman 1994, Kitamura \& Ichimaru 1995,
Brown \& Sawyer 1997)

\begin{equation}
<j,k>^*=f_{scr}(Z_j,Z_k,\rho,T,Y_i) <j,k>. 
\end{equation}

This screening factor is dependent on the charge of the involved particles,
the density, temperature, and the composition of the plasma. Here $Y_i$
denotes the abundance of nucleus $i$ defined by $Y_i=n_i/(\rho N_A)$, where
$n_i$ is the number density of nuclei per unit volume and $N_A$ Avogadro's 
number. At high densities and low temperatures screening factors can enhance 
reactions by many orders of magnitude and lead to {\it pycnonuclear ignition}.
In the extreme case of very low temperatures, where reactions are only
possible via ground state oscillations of the nuclei in a Coulomb lattice,
Eq.(1.5) breaks down, because it was derived under the assumption of a
Boltzmann distribution (for recent references see 
Fushiki \& Lamb 1987, Itoh et al. 1990, Schramm \& Koonin 1990,
Ichimaru 1993, Chabrier \& Schatzman 1994, Ichimaru 1996).

When in Eq.(1.2)
particle $k$ is a photon, the relative velocity is always c and quantities
in the integral are not dependent on $d^3n_j$. 
Thus it simplifies to $r_j=\lambda_{j,\gamma} n_j$ and $\lambda_{j,\gamma}$ 
results from an integration of the photodisintegration cross section over a 
Planck distribution for photons of temperature $T$ 

\begin{subeqnarray}
d^3n_\gamma & =  & {{1} \over {\pi^2 (c\hbar)^3}} {{E_\gamma^2}\over
{ {\rm exp}(E_\gamma /kT)-1}} dE_\gamma\\
r_j & = & \lambda_{j,\gamma} (T)n_j= {{\int d^3n_j}\over {\pi ^2 (c \hbar )^3}} 
\int_0 ^\infty {{c \sigma(E_\gamma ) E_{\gamma}^2}
\over {{\rm exp}(E_\gamma /kT) -1}} dE_\gamma. 
\end{subeqnarray}

There is, however, no direct need to evaluate photodisintegration cross 
sections, because, due to detailed balance, they can be expressed by the 
capture cross sections for the inverse reaction $l+m\rightarrow j+\gamma$
(Fowler et al. 1967)

\begin{equation}
\lambda_{j,\gamma} (T)= ({{G_l G_m} \over G_j}) ({{A_l A_m} \over A_j})^{3/2}
({{m_ukT} \over {2\pi \hbar^2}})^{3/2} <l,m> {\rm exp} (-Q_{lm}/kT).
\end{equation}
This expression depends on the reaction Q-value $Q_{lm}$, the temperature $T$,
the inverse reaction rate $<l,m>$, the partition functions
$G(T)=\sum_i (2J_i+1){\rm exp}(-E_i/kT)$  and the mass numbers $A$ 
of the participating nuclei in a thermal bath of temperature $T$. 

A procedure similar to Eq.(1.6) is used for electron captures by nuclei. 
Because the electron is about 2000 times less massive than a nucleon, the
velocity of the nucleus $j$ is negligible in the center of mass system in
comparison to the electron velocity ($\vert \vec v_j- \vec v_e \vert
\approx \vert \vec v_e \vert$). The electron
capture cross section has to be integrated over a Boltzmann, partially
degenerate, or Fermi distribution of electrons, dependent on the astrophysical
conditions. 
The electron capture rates are a function of $T$ and $n_e=Y_e \rho N_A$, the 
electron number density (Fuller, Fowler, \& Newman 1980, 1982, 1985). In a 
neutral, completely
ionized plasma, the electron abundance is equal to the total proton abundance 
in nuclei $Y_e=\sum_i Z_i Y_i$ and

\begin{equation}
r_j=\lambda_{j,e} (T,\rho Y_e)n_j. 
\end{equation}

The same authors generalized this treatment for the capture of positrons,
which are in a thermal equilibrium with photons, electrons, and nuclei.
At high densities ($\rho >10^{12}$gcm$^{-3}$) the size of the neutrino 
scattering cross section on nuclei and electrons ensures that enough 
scattering events
occur to thermalize a neutrino distribution. Then also the inverse
process to electron capture (neutrino capture) can occur and the
neutrino capture rate can be expresses similar to Eqs.(1.6) or (1.8),
integrating over the neutrino distribution (e.g. Fuller \& Meyer 1995).
Also inelastic neutrino scattering on nuclei can be expressed in
this form. The latter can cause particle emission, like in photodisintegrations
(e.g. Woosley et al. 1990, Kolbe et al. 1992, 1993, 1995, Qian et al. 1996). 
It is also possible that a thermal equilibrium among
neutrinos was established at a different location than at the point where the 
reaction occurs. In such a case the neutrino distribution can be characterized 
by a chemical potential and a temperature which is not necessarily equal to 
the local temperature.
Finally, for normal decays, like beta or alpha decays with half-life 
$\tau_{1/2}$, we obtain an equation similar to Eqs.(1.6) or (1.8) with a decay
constant $\lambda_j=\ln 2/\tau_{1/2}$ and

\begin{equation}
r_j=\lambda_j n_j.
\end{equation}

\subsection{Nuclear Reaction Networks}

The time derivative of the number densities of each of the species in an 
astrophysical plasma (at constant density) is governed by the different 
expressions for $r$, the number of reactions per cm$^3$ and sec, as discussed 
above for the different reaction mechanisms which can change nuclear 
abundances

\begin{equation}
({{\partial n_i} \over {\partial t}})_{\rho =const}= 
\sum_j N^i _j r_j + \sum_{j,k} N^i _{j,k} r_{j,k}
+ \sum_{j,k,l} N^i _{j,k,l} r_{j,k,l}.
\end{equation}

The reactions listed on the right hand side of the equation belong to the 
three categories of reactions: (1) decays,
photodisintegrations, electron and positron captures and neutrino
induced reactions ($r_j=\lambda_j n_j$), 
(2) two-particle reactions ($r_{j,k}=<j,k>n_j n_k$), and (3) three-particle 
reactions ($r_{j,k,l}=<j,k,l> n_j n_k n_l$) like the triple-alpha process, 
which can be interpreted as successive captures with an intermediate unstable 
target (see e.g. Nomoto, Thielemann, \& Miyaji 1985, G\"orres, Wiescher,
\& Thielemann 1995).
The individual $N^{i}$'s are
given by:  $N^i_j = N_i$, $N^i_{j,k} = N_i / \prod_{m=1}^{n_m} | N_{j_m} |! $, 
and $N^i_{j,k,l} = N_i / \prod_{m=1}^{n_m} |N_{j_m}|!$. 
The $N_i's$ can be positive or negative numbers and specify how many particles 
of species $i$ are created or destroyed in a reaction. The denominators,
including factorials, run over the $n_m$ different species destroyed in the
reaction and avoid double counting of the number of reactions
when identical particles react with each other (for example in the
$^{12}$C+$^{12}$C or the triple-alpha reaction; for details see Fowler et al.
1967). In order to exclude changes
in the number densities $\dot n_i$, which are only due to expansion or 
contraction of the gas, the nuclear abundances $Y_i =n_i/(\rho N_A)$ were 
introduced. For a nucleus with atomic weight $A_i$, $A_iY_i$ represents
the mass fraction of this nucleus, therefore $\sum A_iY_i=1$.
In terms of nuclear abundances $Y_i$, a reaction network is described by 
the following set of differential equations
\begin{equation}
\dot Y_i = \sum_j N^i _j \lambda_j Y_j + \sum_{j,k} N^i _{j,k} 
\rho N_A <j,k> Y_j Y_k
+ \sum_{j,k,l} N^i _{j,k,l} \rho^2 N_A^2 <j,k,l> Y_j Y_k Y_l. 
\end{equation}

Eq.(1.11) derives directly from Eq.(1.10) when the definition for the $Y_i's$ 
is introduced. This set of differential equations is solved with  
a fully implicit treatment. Then the  stiff set of differential equations 
can be rewritten (see e.g. Press et al. 1986, \S 15.6) 
as difference equations of the form 
$\Delta Y_i/\Delta t=f_i(Y_j(t+\Delta t))$, where
$Y_i(t+\Delta t)=Y_i(t)+\Delta Y_i$. 
In this treatment, all quantities on 
the right hand side are evaluated at time $t+\Delta t$. This results in a 
set of non-linear equations for the new abundances $Y_i(t+\Delta
t)$, which can be solved using a multi-dimensional Newton-Raphson iteration
procedure.
The total energy generation per gram, due to
nuclear reactions in a time step $\Delta t$ which changed the abundances
by $\Delta Y_i$, is expressed in terms of the mass excess $M_{ex,i}c^2$
of the participating nuclei (Audi \& Wapstra 1995)

\begin{subeqnarray}
\Delta \epsilon & = & - \sum_i \Delta Y_i N_A M_{ex,i}c^2 \\
\dot \epsilon  & = & - \sum_i \dot Y_i N_A M_{ex,i}c^2. 
\end{subeqnarray}

As noted above, the important ingredients to nucleosynthesis calculations
are decay half-lives, electron and positron capture rates, photodisintegrations,
neutrino induced reaction rates,
and strong interaction cross sections. Beta-decay half-lives for
unstable nuclei have been predicted by Takahashi, Yamada, \& Kondo (1973),
Klapdor, Metzinger, \& Oda (1984), Takahashi \& Yokoi (1987, also
including temperature effects) and more recently with improved quasi 
particle RPA calculations (Staudt et al.~1989, 1990, M\"oller 
\& Randrup 1990, Hirsch et al.~1992, Pfeiffer \& Kratz 1996, M\"oller, Nix, 
\& Kratz 1997, Borzov 1996, 1997). Electron and positron capture calculations 
have been performed 
by Fuller, Fowler, \& Newman (1980, 1982, 1985) for a large variety of 
nuclei with mass numbers between A=20 and A=60.
For revisions see also Takahara et al. (1989) 
and for heavier nuclei Aufderheide et al.~(1994), Sutaria, Sheikh, \& Ray 
(1997). 
Rates for inelastic neutrino scattering have been presented by Woosley et al. 
(1990) and Kolbe et al. (1992, 1993, 1995). 
Photodisintegration rates can
be calculated via detailed balance from the reverse capture rates.
Experimental nuclear rates for light nuclei have been discussed in detail in
the reviews by Rolfs, Trautvetter, \& Rodney (1987), 
Filippone (1987), the book by Rolfs \& Rodney (1988), the recent
review on 40 years after B$^2$FH by Wallerstein et al. (1997), and 
the NuPECC report on nuclear and particle astrophysics (Baraffe et al. 1997).
The most recent experimental charged particle rate compilations are the ones 
by Caughlan \&
Fowler (1988) and Arnould et al. (1997). 
Experimental neutron capture cross sections are summarized by Bao \&
K\"appeler (1987, 1997), Beer, Voss, \& Winters (1992), and Wisshak et 
al.~(1997). Rates for unstable (light) nuclei 
are given by Malaney \& Fowler (1988, 1989), Wiescher et al.~(1986, 1987, 
1988ab, 1989ab, 1990), Thomas et al.~(1993,1994),
van Wormer et al. (1994), Rauscher et al. (1994), and Schatz et al. (1997).
For additional information see the article by M. Wiescher (this volume).
For the vast number of medium and
heavy nuclei which exhibit a high density of excited states at capture energies,
Hauser-Feshbach (statistical model) calculations are applicable. The most 
recent compilations were provided by Holmes et al. (1975), Woosley et al. 
(1978), and Thielemann, Arnould, \& Truran (1987, for a detailed discussion of 
the methods involved and neutron capture cross sections for heavy unstable
nuclei see also section 3.4 and the appendix in Cowan, Thielemann, Truran 1991).
Improvements in level densities (Rauscher, Thielemann, \& Kratz 1997),
alpha potentials, and the consistent treatment of isospin mixing will
lead to the next generation of theoretical rate predictions (Rauscher et al.
1998). Some of it will be discussed in the following section.

\section{Theoretical Cross Section Predictions}

Explosive nuclear burning in astrophysical environments produces unstable 
nuclei, which again can be targets for subsequent reactions. In addition,
it involves a
very large number of stable nuclei, which are not fully explored
by experiments. Thus, it is  necessary to be able to predict reaction cross 
sections and thermonuclear rates with the aid of theoretical models.
Explosive burning in supernovae involves in general intermediate mass
and heavy nuclei. Due to a large nucleon number they have intrinsically
a high density of excited states. A high level density in the 
compound nucleus at the appropriate excitation energy allows to
make use of the statistical model approach for compound nuclear 
reactions [e.g.~\cite{hausfesh52,mahwei79}, Gadioli \& Hodgson (1992)]
which averages over resonances. Here, we want to present recent results
obtained within this approach and outline in a clear way, where in the 
nuclear chart and for which environment temperatures its 
application is valid.  
It is often colloquially termed that the statistical model is 
only applicable for intermediate and heavy nuclei. However, the only necessary 
condition for its application is a large number of resonances
at the appropriate bombarding energies, so that the cross section can be
described by an average over resonances. This can in specific cases be valid for
light nuclei and on the other hand not be valid for intermediate mass nuclei
near magic numbers. 

In astrophysical applications usually different aspects are emphasized than 
in pure nuclear physics investigations. Many of
the latter in this long and well established field were focused on specific
reactions, where all or most "ingredients", like optical potentials for
particle and alpha transmission coefficients, level densities, resonance
energies and widths of giant resonances to be implementated in predicting
E1 and M1 gamma-transitions, were deduced from experiments. This of course,
as long as the statistical model prerequisites are met, will produce highly
accurate cross sections.
For the majority of nuclei in astrophysical applications such information
is not available. The real challenge is thus not the well established
statistical model, but rather to provide all these necessary ingredients
in as reliable a way as possible, also for nuclei where none of such 
informations are available. In addition, these approaches should be on a
similar level as e.g. mass models, where the investigation of hundreds
or thousands of nuclei is possible with managable computational effort,
which is not always the case for fully microscopic calculations.

The statistical model approach has been employed
in calculations of thermonuclear reaction rates for astrophysical 
purposes by many researchers [\cite{truran66},
Michaud \& Fowler (1970, 1972), \cite{truran72}], who in the beginning only
made use of ground state properties. Later, the
importance of excited states of the target was pointed out by~\cite{A73}. 
The compilations by~\cite{holmes76,woosley78,thielemann87}, and 
Cowan, Thielemann \& Truran (1991) are presently the ones utilized
in large scale applications in all subfields of nuclear astrophysics,
when experimental information is unavailable.
Existing global optical potentials, mass models to predict Q-values,
deformations etc., but also the ingredients to describe giant resonance
properties have been quite successful in the past [see e.g. the review
by Cowan et al.~(1991)]. 

Besides possibly necessary improvements in global alpha potentials (see
Mohr et al.~1997), the major remaining uncertainty
in all existing calculations stems from the prediction of nuclear level
densities, which in earlier calculations gave uncertainties even beyond
a factor of 10 at the neutron separation energy for~\cite{gilcam65},
about a factor of 8 for~\cite{woosley78}, and a factor of 5 even in
the most recent calculations [e.g.~\cite{thielemann87}; see 
Fig.3.16 in~\cite{cowtt91}]. In nuclear reactions the 
transitions to lower lying states dominate due to the strong energy
dependence. Because the deviations are 
usually not as high yet at low excitation energies, the typical cross 
section uncertainties amounted to a smaller factor of 2--3.

The implementation of a novel treatment of
level density descriptions~\cite{ilji92,igna79}, where
the level density parameter is energy dependent and shell effects vanish at
high excitation energies, improves the level density accuracy. 
This is still a phenomenological approach, making use of a back-shifted
Fermi-gas model rather than a combinatorial approach based on microscopic 
single-particle levels. But it is the first one leading to a
reduction of the average cross section uncertainty
to a factor of about 1.4, i.e.~an average deviation of about 40\% 
from experiments,
when only employing global predictions for all input parameters
and no specific experimental knowledge. 

\subsection
{Thermonuclear Rates from Statistical Model Calculations} 

A high level density in the compound nucleus permits to use averaged
transmission coefficients $T$, which do not reflect a resonance behavior,
but rather describe absorption via an imaginary part in the (optical)
nucleon-nucleus potential as described in Mahaux \& Weidenm\"uller (1979).
This leads to the well known expression

\begin{eqnarray}
\sigma^{\mu \nu}_{i} (j,o;E_{ij})& = &
{{\pi \hbar^2 /(2 \mu_{ij} E_{ij})} \over {(2J^\mu_i+1)(2J_j+1)}} 
\nonumber \\
\label{cslab}
 & & \times \sum_{J,\pi} (2J+1){{T^\mu_j (E,J,\pi ,E^\mu_i,J^\mu_i,
\pi^\mu_i) T^\nu_o (E,J,\pi,E^\nu_m,J^\nu_m,\pi^\nu_m)} \over
{T_{tot} (E,J,\pi)}}
\end{eqnarray}
\par\noindent
for the reaction $i^\mu (j,o) m^\nu$ from the target
state $i^{\mu}$ to the exited state $m^{\nu}$ of the final nucleus, with a
center of mass energy E$_{ij}$ and reduced mass $\mu _{ij}$. $J$ denotes the
spin, $E$ the corresponding excitation energy in the compound nucleus,
and $\pi$ the parity of excited states.
When these properties are used  without subscripts they describe the compound
nucleus, subscripts refer to states of the participating nuclei in the
reaction $i^\mu (j,o) m^\nu$
and superscripts indicate the specific excited states. 
Experiments measure $\sum_{\nu} \sigma_{i} ^{0\nu} (j,o;E_{ij})$,
summed over all excited states of
the final nucleus, with the target in the ground state. Target states $\mu$ in
an astrophysical plasma are thermally populated and the astrophysical cross
section $\sigma^*_{i}(j,o)$ is given by
\begin{equation}
\label{csstar}
\sigma^*_{i} (j,o;E_{ij}) = {\sum_\mu (2J^\mu_i+1) \exp(-E^\mu_i /kT)
\sum_\nu \sigma^{\mu \nu}_{i}(j,o;E_{ij}) \over \sum_\mu (2J^\mu_i+1)
 \exp(-E^\mu_i/kT)}\quad.
\end{equation}
The summation over $\nu$ replaces $T_o^{\nu}(E,J,\pi)$ in Eq.(\ref{cslab}) by
the total transmission coefficient
\begin{eqnarray}
T_o (E,J,\pi) & = &\sum^{\nu_m}_{\nu =0} 
T^\nu_o(E,J,\pi,E^\nu_m,J^\nu_m, \pi^\nu_m) \nonumber \\
\label{tot}
& &+ \int\limits_{E^{\nu_m}_m}^{E-S_{m,o}} \sum_{J_m,\pi_m}
T_o(E,J,\pi,E_m,J_m,\pi_m)\rho(E_m,J_m,\pi_m) dE_m\quad.
\end{eqnarray}
Here $S_{m,o}$ is the channel separation energy, and the summation over
excited 
states above the highest experimentally
known state $\nu_m$ is changed to an integration over the level density
$\rho$.
The summation over target states $\mu$ in Eq.(\ref{csstar}) has to be 
generalized accordingly. 

In addition to the ingredients required for Eq.(\ref{cslab}), like the
transmission coefficients for particles and photons, 
width fluctuation corrections $W(j,o,J,\pi)$ have to be
employed. They define the correlation factors with which all
partial channels for an incoming particle $j$ and outgoing particle $o$,
passing through the excited state $(E,J,\pi)$, have to be multiplied.
This takes into account that the decay of the state is not fully
statistical, but some memory of the way of formation is retained and
influences the available decay choices. The major effect is elastic
scattering, the incoming particle can be immediately re-emitted before
the nucleus equilibrates. Once the particle is absorbed and not
re-emitted in the very first (pre-compound) step, the equilibration is
very likely. This corresponds to enhancing the elastic channel by a
factor $W_j$. In order to conserve the total cross
section, the individual transmission coefficients in the outgoing
channels have to be renormalized to $T_j^\prime $. The total 
cross section is proportional to $T_j$ and, when summing over the elastic 
channel ($W_jT_j^\prime$) and all outgoing channels 
($T^\prime_{tot}-T^\prime_j$), one obtains the condition 
$T_j$=$T_j^\prime (W_jT_j^\prime/T^\prime_{tot})+T^\prime_j(T^\prime_{tot}
-T^\prime_j)/T^\prime_{tot}$. We can (almost) solve for $T^\prime_j$
\begin{equation}
\label{widthcorr}
T^\prime_j={T_j\over 1+ T^\prime_j(W_j-1)/T^\prime_{tot}}\quad.
\end{equation}
This requires an iterative solution for $T^\prime$ (starting in the
first iteration with $T_j$ and $T_{tot}$), which converges fast.
The enhancement factor $W_j$ has to be known in order to apply
Eq.(\ref{widthcorr}). A general expression 
in closed form was derived by~\cite{verbaatschot86}, but is computationally 
expensive to
use. A fit to results from Monte Carlo calculations by~\cite{tepel74} gave
\begin{equation}
\label{newcorr}
W_j=1+{2\over 1+ T_j^{1/2}}\quad.
\end{equation}

For a general discussion of approximation methods see 
\cite{gaho92} and \cite{ezplu93}.
Eqs.(\ref{widthcorr}) and (\ref{newcorr}) redefine the transmission
coefficients of Eq.(\ref{cslab}) in such a manner that the total width is
redistributed by enhancing the elastic channel and weak channels over
the dominant one. Cross sections near threshold energies of new channel 
openings, where very different channel strengths exist, can only be
described correctly when taking width fluctuation corrections into
account. Of the thermonuclear rates presently available
in the literature, only those by~\cite{thielemann87} and Cowan et al.~(1991) 
included this
effect, but their level density treatment still contained large uncertainties. 
The width fluctuation
corrections of~\cite{tepel74} are only an approximation to the
correct treatment. However,~\cite{thomasea86} showed that
they are quite adequate.

The important ingredients of statistical model calculations as indicated in
Eqs.(\ref{cslab}) through (\ref{tot})
are the particle and gamma-transmission coefficients $T$ and
the level density of excited states $\rho$. Therefore, the reliability of  
such calculations is determined by the accuracy with which these components 
can be evaluated (often for unstable nuclei). In the following we want to 
discuss
the methods utilized to estimate these quantities and recent improvements.

\subsection{Transmission Coefficients}
The transition from an excited state in the compound nucleus $(E,J,\pi)$
to the state $(E^\mu_i,J^\mu_i,\pi^\mu_i)$ in nucleus $i$ via the emission of
a particle $j$ is given by a summation over all quantum mechanically allowed
partial waves
\begin{equation}
T^\mu_j (E,J,\pi,E^\mu_i,J^\mu_i,\pi^\mu_i) =
\sum_{l=\vert J-s \vert}^{J+s}
\sum_{s=\vert J^\mu_i -J_j \vert}^{J^\mu_i + J_j}
T_{j_{ls}} (E^\mu_{ij}). 
\end{equation}
Here the angular momentum $\vec l$ and the channel spin $\vec s =\vec J_j+
\vec J^\mu_i$ couple to $\vec J = \vec l +\vec s$. The transition energy
in channel $j$ is $E^\mu_{ij}$=$E-S_j-E^\mu_i$.

The individual particle transmission
coefficients $T_l$ are calculated by solving the Schr\"o\-dinger equation
with an optical potential for the particle-nucleus interaction. All early
studies of thermonuclear reaction rates by~\cite{truran66,michfow72,A73,
truran72,holmes76}, and~\cite{woosley78}
employed optical square well
potentials and made use of the black nucleus approximation. 
Thielemann et al.~(1987) employed the optical potential for neutrons and
protons given by Jeukenne, Lejeune, \& Mahaux (1977), based on microscopic
infinite nuclear matter calculations for a given density,
applied with a local density approximation. 
It includes corrections of the imaginary part by~\cite{fantoni81} 
and~\cite{mahaux82}.
The resulting s-wave neutron strength
function $<\Gamma^o/D> \vert_{\rm 1eV}=(1/2\pi )T_{n(l=0)}(\rm 1eV)$
is shown and discussed in~\cite{thielemann83} and~\cite{cowtt91},
where several phenomenological optical potentials of the 
Woods-Saxon type and the equivalent square well potential used in earlier 
astrophysical applications are compared.
The purely theoretical approach gives the best fit. It is also expected to 
have the most reliable extrapolation properties for unstable nuclei.
We show here in Fig.~1 the ratio of the s-wave strength functions
for the Jeukenne, Lejeune, \& Mahaux potential over the black nucleus,
equivalent square well approach for different energies.
A general overview on different approaches can be found in~\cite{varner91}.

Deformed nuclei were treated in a very simplified way
by using an effective spherical potential of equal
volume, based on averaging the deformed potential over all possible
angles between the incoming particle and the orientation of the deformed
nucleus.
In most earlier compilations alpha particles were also treated 
by square well optical potentials. \cite{thielemann87} employed a 
phenomenological 
Woods-Saxon potential by~\cite{Mann78}, based on extensive data 
from~\cite{macsat66}.
For future use, for alpha particles and heavier projectiles,
it is clear that
the best results can probably be obtained with folding potentials 
[e.g.~\cite{satchlov79,chaud85,oberhu96}, and Mohr et al.~(1997)].

\begin{figure}
\vspace{-2.5cm}
\epsfig{file=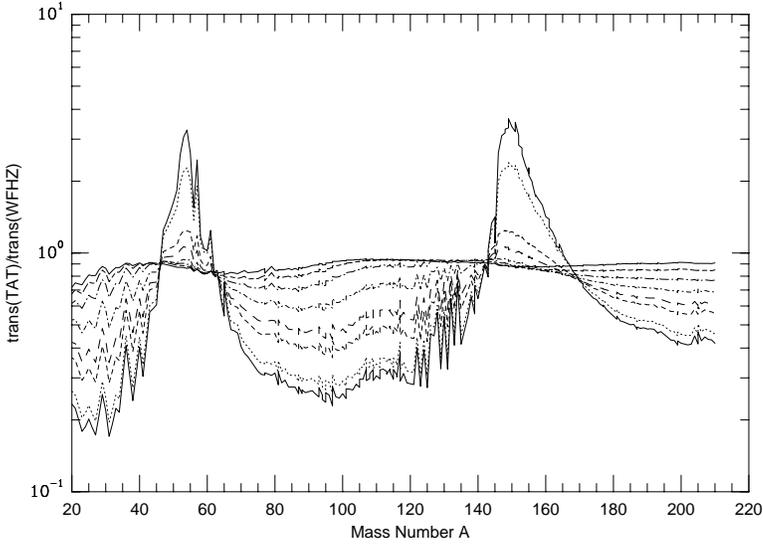,width=9cm,angle=0}
\vspace{-0.3cm}
\caption{\label{s_trans_n}Ratios of transmission functions obtained with the 
\cite{jeukenne77}
potential and the equivalent square well description of~\cite{woosley78}. 
Shown are the ratios for s-wave neutrons. Different
line styles denote different energies:
0.01 MeV (solid), 0.1 MeV
(dotted), 1.0 MeV (short dashes), 2.0 MeV (long dashes), 5.0 MeV (dot --
short dash), 10.0 MeV (dot -- long dash), 15.0 MeV (short dash -- long dash),
20.0 MeV (solid).}
\end{figure}

The gamma-transmission coefficients have to include 
the dominant gamma-transitions (E1 and M1) 
in the calculation of the total photon width.
The smaller, and therefore less important,
M1 transitions have usually been treated with the simple single particle
approach $T \propto E^3$ of~\cite{BW52}, as also discussed in
\cite{holmes76}. The E1 transitions
are usually calculated on the basis of the Lorentzian representation of the
Giant Dipole Resonance (GDR). Within this model, the E1 transmission
coefficient
for the transition emitting a photon of energy $E_{\gamma}$ in a nucleus
$^A_N Z$ is given by
\begin{equation}
\label{gamtrans}
T_{E1}(E_{\gamma}) = {8 \over 3} {NZ \over A} {e^2 \over \hbar c}
{ {1+\chi}
\over mc^2} \sum_{i=1}^2 {i \over 3} { {\Gamma_{G,i} E^4_\gamma} \over
{(E_\gamma^2 -E^2_{G,i})^2 + \Gamma^2_{G,i} E^2_\gamma}}\quad.
\end{equation}
Here $\chi(=0.2)$ accounts for the neutron-proton exchange 
contribution as discussed in Lipparini \& Stringari (1989), and the
summation over $i$ includes two terms which correspond to the split of the
GDR in statically deformed nuclei, with oscillations along (i=1) and
perpendicular (i=2) to the axis of rotational symmetry. Many microscopic
and macroscopic models have been devoted to the calculation of the GDR
energies ($E_{G}$) and widths ($\Gamma_G$).
Analytical fits as a function of $A$ and $Z$ were also used, e.g. 
in~\cite{holmes76} and~\cite{woosley78}.
\cite{thielemann87} employed the (hydrodynamic) droplet model approach 
by~\cite{myersea77} for $E_G$,
which gives an excellent fit to the GDR energies and
can also predict the split of the resonance for deformed nuclei, when
making use of the deformation, calculated within the droplet model.
In that case, the two resonance energies are related to the
mean value calculated by the relations
$E_{G,1}+2E_{G,2}=3E_G$, $E_{G,2}/E_{G,1}=0.911 \eta +0.089$
of~\cite{danos58}.
$\eta$ is the ratio of the diameter along the nuclear
symmetry axis to the diameter perpendicular to it, and can be
obtained from the experimentally known deformation or mass model
predictions.

Cowan et al.~(1991) also give a detailed description of the 
microscopic-macroscopic
approach utilized to calculate $\Gamma_G$, based on dissipation and the
coupling to quadrupole surface vibrations. This is the method applied to
predict the gamma-transmission coefficients for the cross section
determinations shown in the following.

\subsection{Level Densities}
\label{fgas}

While the method as such is well seasoned,
considerable effort has been put into the improvement of the input for 
statistical Hauser-Feshbach models. However, 
the nuclear level density has given rise to the largest uncertainties in 
cross section determinations of~\cite{holmes76,thielemann87,thielemann88}, and 
\cite{cowtt91}. 
For large scale astrophysical applications it is also necessary to not
only find reliable methods for level density predictions, 
but also computationally feasible ones.

Such a model 
is the non-interacting Fermi-gas model by~\cite{bethe36}. Most statistical 
model calculations use the back-shifted Fermi-gas 
description of Gilbert \& Cameron (1965). 
More sophisticated Monte Carlo shell model calculations, e.g. by~\cite{dean95}, 
as well as combinatorial approaches [see e.g.~\cite{paar}], have
shown excellent agreement with this phenomenological approach and justified
the application of the Fermi-gas description at and above the neutron
separation energy.
Rauscher, Thielemann, \& Kratz (1997) applied
an energy-dependent level density parameter $a$ together with
microscopic corrections from nuclear mass models, which leads
to improved fits in the mass range $20\le A \le 245$.

The back-shifted Fermi-gas description of~\cite{gilcam65} 
assumes an even 
distribution of odd and even parities [however, see e.g.~\cite{pich94} for 
doubts on the validity of this assumption at energies of astrophysical 
interest]

\begin{equation}
\rho(U,J,\pi)={1 \over 2} {\cal F}(U,J) \rho(U) ,
\end{equation}
with
\begin{eqnarray}
\rho(U)={1 \over \sqrt{2\pi} \sigma}{\sqrt{\pi} \over
12a^{1/4}}{\exp(2\sqrt{aU}) \over U^{5/4}}\ ,\qquad
{\cal F}(U,J)={2J+1 \over 2\sigma^2} \exp\left({-J(J+1) \over
2\sigma^2}\right) \\
\sigma^2={\Theta_{\mathrm{rigid}} \over \hbar^2} \sqrt{U \over a}\ ,\qquad
\Theta_{\mathrm{rigid}}={2 \over 5}m_{\mathrm{u}}AR^2\ ,\qquad
U=E-\delta\quad. \nonumber
\end{eqnarray}
The spin dependence ${\cal F}$ is determined by the spin cut-off parameter 
$\sigma$. Thus, the level density is dependent on only two parameters: 
the level density parameter $a$ and the backshift $\delta$, which 
determines the energy of the first excited state.

Within this framework, the quality of level density predictions depends 
on the reliability of systematic estimates of $a$ and $\delta$. The 
first compilation for a large number of nuclei was provided 
by~\cite{gilcam65}. They found that the backshift $\delta$ is well 
reproduced by experimental pairing corrections (Cameron \& Elkin 1965). 
They also were the first 
to identify an empirical correlation with experimental shell corrections 
$S(Z,N)$
\begin{equation}
\label{aovera}
{a \over A}=c_0+c_1S(Z,N),
\end{equation}
where $S(Z,N)$ is negative near closed shells. 
The back-shifted Fermi-gas approach diverges for $U=0$ (i.e.\ $E=\delta$, if
$\delta$ is a positive backshift). In order to obtain the correct behavior at
very low excitation energies, the Fermi-gas description
can be combined with the constant temperature formula [\cite{gilcam65},
\cite{gaho92} and references therein]
\begin{equation}
\label{ctemp}
\rho(U) \propto {\exp(U/T) \over T}\quad.
\end{equation}
The two formulations are matched by a tangential fit determining $T$.
There have been a number of compilations for $a$ and $\delta$, or $T$,
based on experimental level densities, as 
e.g.~the ones by von Egidy et al. (1986,1988). 
An improved approach has to consider the energy dependence of the shell 
effects, which are known to vanish at high excitation energies, see
e.g.~\cite{ilji92}.
Although, for astrophysical purposes only energies close to the particle 
separation thresholds have to be considered, an energy dependence can 
lead to a considerable improvement of the global fit. This is especially 
true for strongly bound nuclei close to magic numbers.

An excitation-energy dependent description was initially proposed 
by Ignatyuk et al.~(1975) and~\cite{igna79}
for the level density parameter $a$
\begin{subeqnarray}
\label{endepa}
a(U,Z,N)& = &\tilde{a}(A)\left[1+C(Z,N){f(U) \over 
U}\right]\\
\tilde{a}(A)& = &\alpha A+\beta A^{2/3}\\
f(U)& = & 1-\exp(-\gamma U).
\end{subeqnarray}
The values of the free parameters $\alpha$, $\beta$ and $\gamma$ are
determined by fitting to experimental level density data available over
the whole nuclear chart.

The shape of the function $f(U)$ permits the two extremes: (i) for small
excitation energies the original form of Eq.(\ref{aovera}) 
$a/A=\alpha+\alpha\gamma C(Z,N)$ is retained
with $S(Z,N)$ being replaced by $C(Z,N)$, (ii)
for high excitation energies $a$/$A$ approaches the continuum value 
$\alpha$, obtained for infinite nuclear matter. In both cases we neglected
$\beta$, which is realistic as discussed below.
Previous attempts to find a global description of level densities used 
shell corrections $S$ derived from comparison of liquid-drop 
masses with experiment ($S\equiv M_{\mathrm{exp}}-M_{\mathrm{LD}}$) or 
the ``empirical'' shell corrections $S(Z,N)$ given by~\cite{gilcam65}.
A problem 
connected with the use of liquid-drop masses arises from the fact that 
there are different liquid-drop model parametrizations available in the
literature which produce quite different values for $S$, as shown
in Mengoni \& Nakayama (1994).

However, in addition, the meaning of the correction parameter inserted into
the level density formula Eq.(\ref{endepa}) has to be reconsidered. 
The fact that nuclei approach a spherical shape at high excitation energies
(temperatures) has to be included.
Therefore, the correction parameter $C$ should 
describe properties of a nucleus differing from the {\it spherical} macroscopic 
energy and contain those terms which are finite for low and vanishing at 
higher excitation energies. The latter requirement is mimicked by the form
of Eq.(\ref{endepa}).
Therefore, the parameter $C(Z,N)$ should rather be identified with 
the so-called ``microscopic'' correction $E_{\mathrm{mic}}$ than with the
shell correction. The mass of 
a nucleus with deformation $\epsilon$ can then be written in two ways
\begin{subeqnarray}
\label{emic}
M(\epsilon)& = &E_{\mathrm{mic}}(\epsilon)+E_{\mathrm{mac}}
(\mathrm{spherical})\\
M(\epsilon)& = &E_{\mathrm{mac}}(\epsilon)+E_{\mathrm{s+p}}
(\epsilon),
\end{subeqnarray}
with $E_{\mathrm{s+p}}$ being the shell-plus-pairing correction.
This confusion about the term ``microscopic correction'', being 
sometimes used in an ambiguous way, is also pointed out 
in~\cite{moeller95}. 
The above mentioned ambiguity 
follows from the inclusion of deformation-dependent effects
into the macroscopic part of the mass formula.

Another important ingredient is
the pairing gap $\Delta$, related to the backshift $\delta$. 
Instead of assuming constant pairing as in \cite{Reisdorf}
or an often applied fixed dependence on the mass
number via e.g. $\pm 12/\sqrt{A}$, 
the pairing gap $\Delta$ can be determined from differences in the binding
energies (or mass differences, respectively) of
neighboring nuclei.
Thus, similar to Ring \& Schuck (1980),~\cite{wang92} obtained
for the neutron pairing gap $\Delta_{\mathrm{n}}$
\begin{equation}
\label{pair}
\Delta_{\mathrm{n}}(Z,N)={1 \over 2} \left[ 
M(Z,N-1)+M(Z,N+1)-2M(Z,N)\right],
\end{equation}
where $M(Z,N)$ is the ground state mass excess of the nucleus $(Z,N)$.
Similarly, the proton pairing gap $\Delta_{\mathrm{p}}$ can be 
calculated.

\subsection{Results}
\noindent
In our study we utilized the microscopic corrections of the recent mass 
formula by M\"oller et al.~(1995), calculated with the Finite Range 
Droplet Model FRDM
(using a folded Yukawa shell model with Lipkin-Nogami pairing)
in order to determine the parameter $C(Z,N)$=$E_{\mathrm{mic}}$. The 
backshift $\delta$ was calculated by 
setting $\delta(Z,N)$=1/2$\{\Delta_{\mathrm{n}}(Z,N)+\Delta_{\mathrm{p}}(Z,N)\}$
and using Eq.(\ref{pair}).
The parameters $\alpha$, $\beta$, and $\gamma$ were obtained from a fit to 
experimental data for s-wave neutron resonance spacings of 272 nuclei at 
the neutron separation energy. The 
data were taken from the compilation by~\cite{ilji92}.
Similar investigations were recently undertaken by~\cite{mengna94},
who made, however, use of a slightly different description of the energy 
dependence of $a$ and of different pairing gaps.

\begin{figure}
\epsfig{file=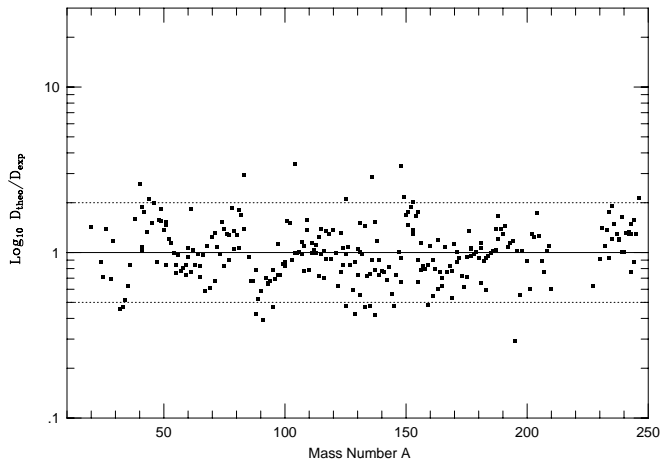,angle=90,width=10cm}
\caption{\label{figrat}Ratio of predicted to 
experimental~\protect{\cite{ilji92}} level 
densities at the neutron separation energy. The deviation is less than a 
factor of 2 (dotted lines) for the majority of the considered nuclei.}
\end{figure}

As a quantitative overall estimate of the agreement between calculations
and experiments, one usually quotes the ratio
\begin{equation}
g\equiv \left< {\rho_{\mathrm{calc}} \over \rho_{\mathrm{exp}}}\right> =
\exp \left[{1 \over n} \sum_{i=1}^{n}\left( \ln {\rho_{\mathrm{calc}}^i
\over \rho_{\mathrm{exp}}^i} \right)^2 \right]^{1/2},
\end{equation}
with $n$ being the number of nuclei for which level densities 
$\rho$ are experimentally known.
As best fit 
we obtain an averaged ratio $g=1.48$ with the parameter 
values $\alpha=0.1337$, $\beta=-0.06571$, $\gamma=0.04884$. 
This corresponds to $a/A=\alpha=0.134$ for infinite nuclear matter, which is
approached for high excitation energies.
The ratios of 
experimental to predicted level densities (i.e. theoretical to experimental
level spacings $D$) for the nuclei considered are
shown in Fig.~\ref{figrat}. As can be seen, for the majority of nuclei 
the absolute deviation is less than a factor of 2. This is a 
satisfactory improvement over theoretical level densities used in previous 
astrophysical cross section calculations, where deviations of a 
factor 3--4, or even in excess of a factor of 
10 were found [for details see ~\cite{cowtt91}]. Such a direct comparison as in 
Fig. 2 was rarely shown in earlier work. In most cases the level
density parameter $a$, entering exponentially into the level density, was
displayed.

Although we quoted the value of the parameter $\beta$ above, it
is small in comparison to $\alpha$ and can be set to zero without considerable 
increase in the
obtained deviation. Therefore, actually only two
parameters are needed for the level density description.
Rauscher, Thielemann, \& Kratz (1997) also tested the sensitivity with respect 
to the employed mass formula.
This phenomenological approach, still in the framework of the back-shifted
Fermi gas model, but with an energy dependent level density parameter, based
on microscopic corrections of nuclear mass models, gives better
results than a recent BCS approach by~\cite{goriely96}, which tried to
implement level spacings from the ETFSI model (Extended Thomas-Fermi with 
Strutinski Integral, Aboussir et al. 1995) in a consistent combinatorial
fashion.

With these improvements, the uncertainty in the level density is now comparable
to uncertainties in optical potentials and gamma transmission coefficients
which enter the determinations of capture cross sections. 
The remaining uncertainty of extrapolations is the one due to the reliability 
of the nuclear structure model applied far from stability  which
provides the microscopic corrections and pairing gaps. We will discuss this in 
more detail in section 5 (see also the contribution by M. Wiescher, 
this volume).

\subsection{Applicability of the Statistical Model}

\noindent
Having a reliable level density description also permits to analyze
when and where the statistical model approach is valid. 
Generally speaking, in order to apply the model correctly, a sufficiently 
large number of levels in the compound nucleus is needed in the relevant 
energy range, which can act as doorway states to the formation of the compound 
nucleus.
In the 
following this is discussed for neutron-, proton- and alpha-induced reactions
with the aid of the level density approach presented above.
This section is intended to be a guide to a meaningful and correct 
application of the statistical model.

The nuclear reaction rate per particle pair at a given stellar temperature
$T$ is determined by folding the reaction cross section with the
Maxwell-Boltzmann (MB) velocity distribution of the projectiles, as displayed
in Eq.(1.4).
Two cases have to be considered, reactions between charged particles and
reactions with neutrons.

\subsubsection{The Effective Energy Window}

\noindent
The nuclear cross section for charged particles is strongely suppressed
at low energies due to the Coulomb barrier.
For particles having energies less than the height of the Coulomb barrier,
the product of the penetration factor and the MB distribution function
at a given
temperature results in the
so-called Gamow peak, in which most of the reactions will
take place. Location and width of the Gamow peak depend 
on the charges
of projectile and target, and on the temperature of the interacting plasma.

When introducing the astrophysical $S$ factor $S(E)=\sigma(E)E\exp(2\pi
\eta)$ (with $\eta$ being the Sommerfeld parameter, describing the s-wave
barrier penetration), one can easily see
the two contributions of the velocity distribution and the penetrability
in the integral
\begin{equation}
< \sigma v >=\left( \frac{8}{\pi \mu} \right) ^{1/2}
\frac{1}{\left( kT \right) ^{3/2}}
\int_0^{\infty} S(E) \exp \left[ - \frac{E}{kT} - \frac{b}{E^{1/2}} 
\right] \quad,
\end{equation}
where the quantity $b=2\pi\eta E^{1/2}=(2 \mu)^{1/2} \pi e^2 Z_j Z_k / 
\hbar$ arises from
the barrier penetrability. Taking the first derivative of the integrand
yields the location $E_0$ of the Gamov peak, and the effective width
$\Delta$ of the energy window can be derived accordingly 
\begin{subeqnarray}
E_0&= &\left( \frac{bkT}{2} \right) ^{2/3}=1.22(Z_j^2 Z_k^2 A T_6^2)^{1/3}
\, \mathrm{keV},\\
\Delta&=&\frac{16E_0 kT}{3}^{1/2}=0.749 (Z_j^2 Z_k^2 A T_6^5)^{1/6}
\, \mathrm{keV},
\end{subeqnarray}
as shown in~\cite{fowler67} and~\cite{rolfs}, 
where the charges $Z_j$, $Z_k$, the reduced mass $A$ of the involved
nuclei in units of $m_u$, and the temperature $T_6$ given in 10$^6$ K, enter. 

In the case of neutron-induced reactions 
the effective energy window has to be derived in a slightly different
way. For s-wave neutrons ($l=0$) the energy window is simply given by the
location and width of the peak of the MB distribution function.
For higher partial waves the penetrability of the centrifugal barrier
shifts the effective energy $E_0$ to higher energies. 
For neutrons with energies less than the height of the
centrifugal barrier this was approximated by \cite{wagoner}
\begin{subeqnarray}
E_0& \approx &0.172 T_9\left( l+{1 \over 2}\right)\quad \mathrm{MeV,}\\
\Delta &\approx &0.194 T_9\left( l+{1 \over 2}\right)^{1/2} \mathrm{MeV.}
\end{subeqnarray}
\noindent
The energy $E_0$ will always be comparatively close to the neutron
separation energy.

\begin{figure}
\vspace{-2.5cm}
\epsfig{file=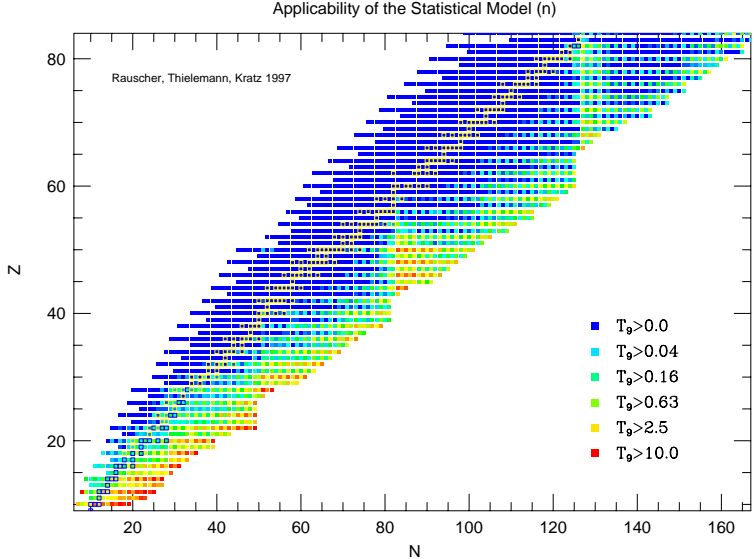,width=9cm,angle=0}
\caption{\label{neutcrit} Stellar temperatures (in 10$^9$ K) for 
which the statistical
model can be used. Plotted is the compound nucleus of the neutron-induced
reaction n+Target. Stable nuclei are marked.}
\end{figure}
\subsubsection{A Criterion for the Application of the Statistical Model}

\noindent
Using the above effective energy windows for charged and neutral
particle reactions, a criterion for the 
applicability can be derived from the level density. 
For a reliable application of the statistical model a sufficient number
of nuclear levels has to be within the energy window, thus contributing
to the reaction rate. For narrow, isolated resonances, the cross sections
(and also the reaction rates) can be represented by a sum over individual
Breit-Wigner terms. The main question is whether the density of resonances 
(i.e. level density) is high enough so that the integral over the sum of 
Breit-Wigner resonances may be approximated by an integral over the
statistcial model expressions of Eq.(\ref{cslab}), which assume that
at any bombarding energy a resonance of any spin and parity is available
[see \cite{wagoner}].

Numerical test calculations have been performed by Rauscher et al. (1997)
in order to find the average number of 
levels per energy window which is sufficient to allow this substitution 
in the specific case of folding over a MB distribution. To achieve 20\% 
accuracy, about 10 levels in total are needed in the effective energy window in
the worst case (non-overlapping, narrow resonances). This relates to a
number of s-wave levels smaller than 3.
Application of the statistical model for a level density which is
not sufficiently large, results in general in
an overestimation of the actual cross section, unless a strong s-wave resonance
is located right in the energy window [see the discussion in~\cite{wormer94}].
Therefore, we will assume in the following a conservative limit of 10 
contributing resonances in the effective energy window for charged and neutral 
particle-induced reactions.

\begin{figure}
\vspace{-2.5cm}
\epsfig{file=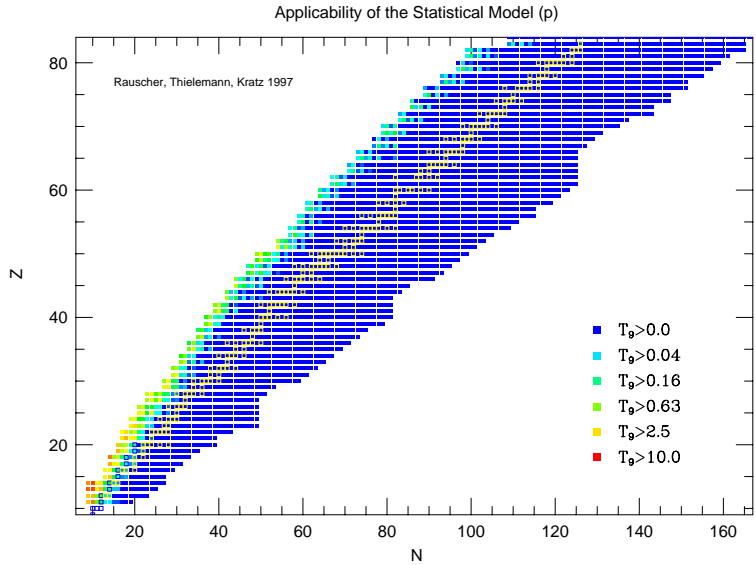,width=9cm,angle=0}
\caption{\label{prot} Stellar temperatures (in 10$^9$) for which 
the statistical 
model can be used. Plotted is the compound nucleus of the proton-induced 
reaction p+Target. Stable nuclei are marked.}
\end{figure}

\begin{figure}
\vspace{-2.5cm}
\epsfig{file=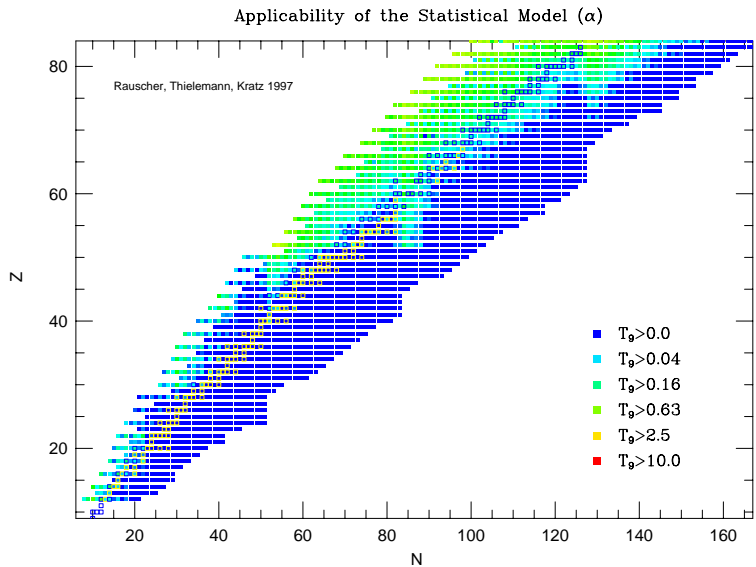,width=9cm,angle=0}
\caption{\label{alph} Stellar temperatures (in 10$^9$) for which the 
statistical model can be
used. Plotted is the compound nucleus of the alpha-induced reaction
alpha+Target. Stable nuclei are marked.}
\end{figure}

To obtain the necessary number of levels within the energy window of width 
$\Delta$ can require a sufficiently high excitation energy, as the level density
increases with energy. This combines with the thermal distribution of 
projectiles to a minimum temperature for the application of the statistical 
model. Those temperatures
are plotted in a logarithmic grey scale in Figs.~\ref{neutcrit}--\ref{alph}.
For neutron-induced reactions Fig.~\ref{neutcrit} applies,
Fig.~\ref{prot} describes proton-induced reactions, and
Fig.~\ref{alph} alpha-induced reactions.
Plotted is always the minimum stellar temperature $T_9$ (in 10$^9$ K) at the 
location of the compound nucleus in the nuclear chart.
It should be noted that the derived temperatures will not change considerably,
even when changing the required level number within a factor of about two,
because of the exponential dependence of the level density on the
excitation energy.

This permits to read directly from the
plot whether the statistical model cross section can be ``trusted'' 
for a specific astrophysical application at a specified temperature or
whether single resonances or other processes (e.g.\ direct reactions) 
have also to be considered. These 
plots can give hints on when it is safe to use the statistical 
model approach and which nuclei have to be treated with special 
attention for specific temperatures. Thus, information on which nuclei 
might be of special interest for an experimental investigation may also 
be extracted.

\section{Nucleosynthesis Processes in Stellar Evolution and Explosions}

Nucleosynthesis calculations can in general be classified into two
categories:
(1) nucleosynthesis during hydrostatic burning stages of stellar evolution
on long timescales and
(2) nucleosynthesis in explosive events (with different initial fuel 
compositions, specific to the event).
In the following we want to discuss shortly reactions of importance
for both conditions and the major burning products.

\subsection{Hydrostatic Burning Stages in Presupernova Evolution} 

The main hydrostatic burning stages and most important reactions are:

\noindent
{\it H-burning:} there are two alternative reaction sequences, the different 
pp-chains which convert $^1$H into
$^4$He, initiated by $^1$H($p,e^+$)$^2$H, and the CNO cycle
which converts $^1$H into $^4$He by a sequence of ($p,\gamma$) and
($p,\alpha$) reactions on C, N, and O isotopes and subsequent beta-decays. 
The CNO isotopes are all transformed into $^{14}$N, due to the fact that the
reaction $^{14}$N($p,\gamma)^{15}$O is the slowest reaction in the
cycle.

\noindent
{\it He-burning:} the main reactions are the triple-alpha reaction 
$^4$He(2$\alpha,\gamma$)$^{12}$C and $^{12}$C($\alpha, \gamma$)$^{16}$O.

\noindent
{\it C-burning:} $^{12}$C($^{12}$C, $\alpha$)$^{20}$Ne and
$^{12}$C($^{12}$C,$p)^{23}$Na. Most of the $^{23}$Na nuclei will react
with the free protons via $^{23}$Na($p,\alpha$)$^{20}$Ne.

\noindent
{\it Ne-Burning:} $^{20}$Ne($\gamma, \alpha$)$^{16}$O, $^{20}$Ne($\alpha ,
\gamma$) $^{24}$Mg  and $^{24}$Mg($\alpha , \gamma$)$^{28}$Si.
It is important that photodisintegrations start to play a role when 
$30kT$$\approx$$Q$ (as a rule of thumb), with $Q$ being the Q-value of a 
capture reaction. For those conditions sufficient photons with energies
$>$$Q$ exist in the high energy tail of the Planck distribution. As
$^{16}$O($\alpha,\gamma)^{20}$Ne has an exceptionally small Q-value of
the order 4~MeV, this relation holds true for $T$$>$$1.5\times 10^9$K,
which is the temperature for (hydrostatic) Ne-burning.

\noindent
{\it O-burning:} $^{16}$O($^{16}$O,$\alpha$)$^{28}$Si,
$^{16}$O($^{16}$O,$p)^{31}$P, and
$^{16}$O($^{16}$O,$n)^{31}$S($\beta^+$)$^{31}$P.
Similar to carbon burning, most of the $^{31}$P is destroyed by a
($p,\alpha $) reaction to $^{28}$Si.

\noindent
{\it Si-burning:} Si-burning is initiated like Ne-burning by 
photodisintegration reactions which then provide the particles for capture 
reactions. It ends in an 
equilibrium abundance distribution around Fe (thermodynamic equilibrium). 
As this includes all kinds of Q-values (on the average
8-10~MeV for capture reactions along the valley of stability), this
translates to temperatures in excess of 3$\times$10$^9$K, being larger
than the temperatures for the onset of Ne-burning. 
In such an equilibrium (also denoted nuclear statistical equilibrium, NSE)
the abundance of each nucleus is only governed by the temperature $T$,
density $\rho$, its nuclear binding energy $B_i$ and partition function
$G_i=\sum_j (2J^i_j+1){\rm exp}(-E^i_j/kT)$

\begin{equation}
\label{nseeq}
Y_i= (\rho N_A)^{A_i-1} {{G_i} \over {2^{A_i}}} A_i^{3/2}
({{2\pi \hbar^2} \over {m_ukT}})^{{3 \over 2} (A_i-1)} {\rm exp}(B_i/kT)
Y_p^{Z_i} Y_n^{N_i},
\end{equation}
while fulfilling mass conservation $\sum_i A_i Y_i=1$ and charge 
conservation $\sum_i Z_i Y_i =Y_e$ (the total number of protons equals 
the net number of electrons, which is usually changed only by weak interactions 
on longer timescales).
This equation is derived from the relation between chemical potentials
(for Maxwell-Boltzmann distributions)
in a thermal equilibrium ($\mu_i=Z_i\mu_p + N_i\mu_n$), where the subscripts
$n$ and $p$ stand for neutrons and protons. Intermediate quasi-equilibrium
stages (QSE), where clusters of neighboring nuclei are in relative equilibrium
via neutron and proton reactions, but different clusters have total abundances
which are offset from their NSE values, are important during the onset
of Si-burning before a full NSE is reached and during the freeze-out
from high temperatures, which will be discussed in section 3.2.

\noindent
{\it s-process:} the slow neutron capture process leads to the build-up of 
heavy elements during core and shell He-burning, 
where through a series of neutron captures and beta-decays, starting on 
existing heavy nuclei around Fe, nuclei up to Pb and Bi can be synthesized. 
The neutrons are provided by a side branch of He-burning,
$^{14}$N($\alpha, \gamma$)$^{18}$F($\beta^+$)$^{18}$O($\alpha, \gamma$)
$^{22}$Ne($\alpha,n)^{25}$Mg. 
An alternative stronger neutron source in He-shell flashes is the reaction
$^{13}$C($\alpha,n)^{16}$O, which requires admixture of hydrogen and the 
production of $^{13}$C via proton capture on $^{12}$C and a subsequent 
beta-decay.

An extensive  overview over the major and minor reaction  sequences
in all burning stages from helium to silicon burning in massive stars is given
in Arnett \& Thielemann (1985), Thielemann \& Arnett (1985), Woosley \&
Weaver (1995), Hix \& Thielemann (1996) and Nomoto et al. (1997)
(see also Arnett 1996 and for the status of experimental rate uncertainties
M. Wiescher, this volume).
For less massive stars which burn at higher densities, i.e. experience
higher electron Fermi energies, electron captures are already important in 
O-burning and lead
to a smaller Y$_e$ or larger neutron excess $\eta =\sum_i (N_i-Z_i)Y_i=1-2Y_e$.
For a general overview of the s-process see
K\"appeler, Beer, \& Wisshak (1989), K\"appeler et al. (1994), Wisshak
et al. (1997), and Gallino \& Busso (1997).

Most reactions in hydrostatic burning stages proceed through 
stable nuclei. This is simply explained by the long timescales involved.
For a 25M$_\odot$ star, which is relatively massive 
and therefore experiences quite short burning phases, 
this still amounts to: H-burning 7$\times$10$^6$ years, He-burning
5$\times$10$^5$ y, C-burning 600 y, Ne-burning 1 y, O-burning 180 days,
Si-burning 1 d. Because 
all these burning stages are long compared to beta-decay half-lives,
with a few exceptions of long-lived unstable nuclei,
nuclei can decay back to stability before undergoing the next reaction.
Examples of such exceptions are the s-process branchings with a competition
between neutron captures and beta-decays of similar timescales (see e.g.
Gallino \& Busso 1997).

\subsection{Explosive Burning}

Many of the hydrostatic burning processes discussed in section 3.1 can
occur also under explosive conditions at much higher temperatures and on
shorter timescales. The major reactions remain still the same in many
cases, but often the beta-decay half-lives of unstable products are
longer than the timescales of the explosive processes under investigation.
This requires in general the additional knowledge of nuclear cross
sections for unstable nuclei.

Extensive calculations of explosive carbon, neon, oxygen, and silicon
burning, appropriate for supernova explosions, have already been performed 
in the late 60s and early 70s with the accuracies possible in those days
and detailed discussions about the expected abundance patterns 
(for a general review see Trimble 1975; Truran 1985). More recent
overviews in the context of stellar models are given by Trimble (1991) and
Arnett (1995).
Besides minor additions of $^{22}$Ne after He-burning
(or nuclei which originate from it in later burning stages, see section 3.1), 
the fuels
for explosive nucleosynthesis consist mainly of alpha-particle nuclei like
$^{12}$C, $^{16}$O, $^{20}$Ne, $^{24}$Mg, or $^{28}$Si. Because the
timescale of explosive processing is very short (a fraction of a second
to several seconds), only few beta-decays can occur during 
explosive nucleosynthesis events, resulting in heavier nuclei, again with 
N$\approx$Z. However, a spread of nuclei around a line of N$=$Z
is involved and many reaction rates for unstable nuclei have to be known. 
Dependent on the temperature, explosive burning produces intermediate to 
heavy nuclei. We will discuss the individual burning processes below.
For the processes discussed in this section, nuclei
within a few mass units from stability are encountered, where
nuclear masses and decay half-lives are known experimentally.

Two processes differ from the above scenario, where either a large
supply of neutrons or protons is available, the r-process and the rp-process,
denoting rapid neutron or proton capture (the latter also termed
explosive hydrogen burning). In such cases, nuclei
close to the neutron and proton drip lines can be prodruced and beta-decay
timescales can be short in comparison to the process timescales.
In this section we will only discuss
the possible connection between explosive Si-burning and the
r-process.

Burning timescales in stellar evolution are dictated by the energy loss
timescales of stellar environments. Processes like hydrogen and helium
burning, where the stellar energy loss is dominated by the photon
luminosity, choose temperatures with energy generation rates 
equal to the radiation losses. For the later burning stages neutrino losses 
play the dominant role among cooling processes and the burning timescales 
are determined by temperatures where neutrino losses are equal to the
energy generation rate (see the long series of investigations by
Itoh and collaborators, e.g. Itoh et al. 1993, 1994, 1996ab). 
Explosive events are determined by hydrodynamics, causing 
different temperatures and timescales for the burning of available fuel.
We can generalize the question by defining a burning timescale according
to Eq.(1.11) for the destruction of the major fuel nuclei $i$

\begin{equation}
\tau_i=|{Y_i \over \dot Y_i}|.
\end{equation}

\begin{figure}
\vspace{1cm}
\epsfig{file=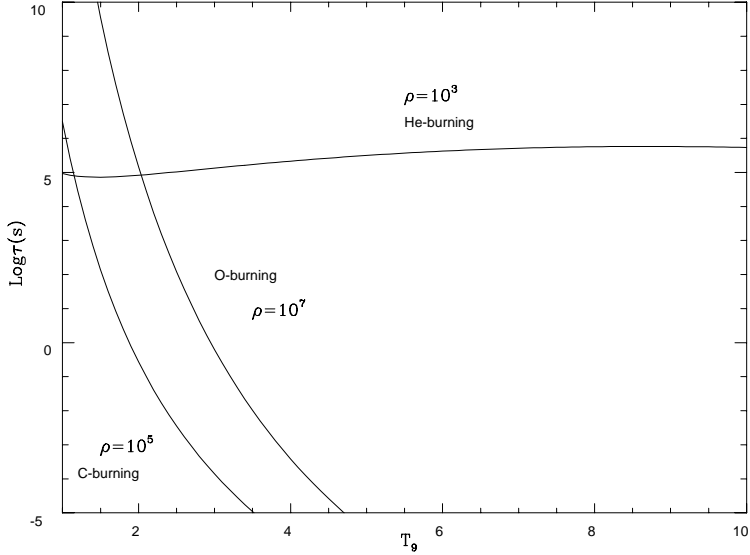,angle=90,width=5cm}
\vspace{1cm}
\caption{Burning timescales for fuel destruction of He-, C-, and O-burning 
as a function of temperature. A 100\% fuel mass fraction was assumed. The 
factor $N^i_{i,i}=N_i/N_i!$ cancels for the destruction of identical
particles by fusion reactions, as $N_i$=2. 
For He-burning the destruction of three identical particles
has to be considered, which changes the leading factor $N^i_{i,i,i}$ to 1/2.
The density-dependent burning timescales are labeled with the
chosen typical density. They scale linearly for C- and O-burning
and quadraticly for He-burning. Notice that the almost constant
He-burning timescale beyond $T_9$=1 has the effect that efficient
destruction on explosive timescales can only be attained for high
densities.
\label{explo1}}
\end{figure}

\begin{figure}
\vspace{1cm}
\epsfig{file=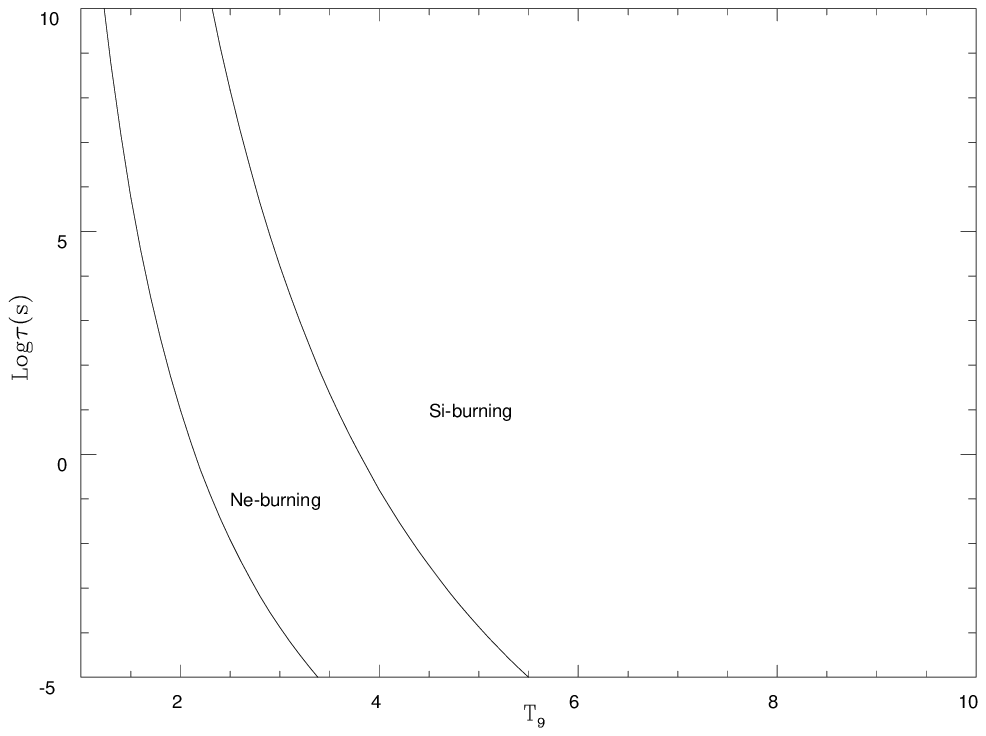,angle=90,width=5cm}
\vspace{1cm}
\caption{Burning timescales for fuel destruction of Ne- and Si-burning 
as a function of temperature. These are burning phases initiated by
photodisintegrations and therefore not density-dependent.
\label{explo2}}
\end{figure}

These timescales for the fuels $i\in$ H, He, C, Ne, O, Si are determinated
by the major distruction reaction. They are in all cases temperature
dependent. Dependent on whether this is (i) a decay or photodisintegration, 
(ii) a two-particle or (iii) a three-particle fusion reaction, they are (i) 
either 
not density dependent or have an inverse (ii) linear or (iii) quadratic
density dependence. Thus, in the burning stages which involve a fusion process,
the density dependence is linear, with the exception of He-burning, where it
is quadratic. Ne- and Si-burning, which are dominated by $(\gamma,\alpha)$
distructions of $^{20}$Ne and $^{28}$Si, have timescales only determined by
the burning temperatures. The temperature dependences are typically
exponential, due to the functional form of the corresponding $N_A<\sigma
v>$ expressions. 
We have plotted these burning timescales as a function of temperature 
(see Figs.~\ref{explo1} and \ref{explo2}), assuming a fuel mass 
fraction of 1. The curves for (also) density dependent burning processes
are labeled with a typical density. 
He-burning has a quadratic density dependence, C- and O-burning
depend linearly on density. 
If we take typical explosive burning timescales to be of the order of
seconds (e.g. in supenovae), we see that one requires temperatures to
burn essential parts of the fuel in excess of 4$\times 10^9$K (Si-burning),
3.3$\times 10^9$K (O-burning), 2.1$\times 10^9$K (Ne-burning), and
1.9$\times 10^9$K (C-burning). Beyond $10^9$K He-burning is
determined by an almost constant burning timescale. We see that
essential destruction on a time scale of 1s is only possible for
densities $\rho$$>$$10^5$g cm$^{-3}$. This is usually not encountered
in He-shells of massive stars. In a similar way explosive H-burning 
is not of relevance for massive stars, but important for explosive burning 
in accreted H-envelopes in binary stellar evolution (these issues are 
discussed by M. Wiescher, this volume).

\subsubsection{Explosive He-Burning}

Explosive He-burning is chararcterized  by the same reactions as
hydrostatic He-burning, producing $^{12}$C and $^{16}$O. 
Fig. \ref{explo1} indicated that even for temperatures beyond $10^9$K 
high densities ($>$$10^5$g cm$^{-3}$) are
required to burn essential amounts of He. 
During the passage of a $10^{51}$erg supernova shockfront
through the He-burning zones of a 25M$_\odot$ star, 
maximum temperatures of only (6-9)$\times 10^8$K
are attained and the amount of He burned is negligible.
However, neutron sources
like $^{22}$Ne($\alpha ,n)^{25}$Mg [or $^{13}$C($\alpha ,n)^{16}$O],
which sustain an s-process neutron flux in hydrostatic burning, 
release a large neutron flux under explosive conditions. This leads to 
partial destruction of $^{22}$Ne and the build-up of $^{25,26}$Mg via
$^{22}$Ne$(\alpha,n)^{25}$Mg($n,\gamma)^{26}$Mg.
Similarly, $^{18}$O and $^{13}$C are destroyed by
alpha-induced reactions. This releases neutrons with
$Y_n$$\approx$2$\times 10^{-9}$ at a density of $\approx$8.3$\times
10^3$g cm$^{-3}$, corresponding to
$n_n$$\approx$10$^{19}$cm$^{-3}$ for about 0.2s, and causes neutron
processing (Truran, Cowan, \& Cameron 1978, Thielemann, 
Arnould, \& Hillebrandt 1979, Thielemann, Metzinger, \& Klapdor 1983, 
Cowan, Cameron, \& Truran 1983). This is, however,
not an r-process (Blake et al. 1981). More detailed calculations for
such mass zones have recently been performed by Howard, Meyer, \&
Clayton (1992).

\subsubsection{Explosive C- and Ne-Burning}

The main burning products of explosive neon burning are $^{16}$O, $^{24}$Mg,
and $^{28}$Si, synthesized via the reaction sequences
$^{20}$Ne($\gamma ,\alpha$)$^{16}$O and
$^{20}$Ne($\alpha ,\gamma$)$^{24}$Mg($\alpha ,\gamma$)$^{28}$Si, 
similar to the hydrostatic case. 
The mass zones in supernovae which undergo explosive neon burning must
have peak temperatures in excess of 2.1$\times 10^9$K.
They undergo a combined version of explosive neon and carbon burning
(see Figs.~\ref{explo1} and \ref{explo2}). Mass zones which experience 
temperatures in excess of
1.9$\times 10^9$K will undergo explosive carbon burning, as long as carbon
fuel is available. This is often not the case in type II supernovae
originating from massive stars. 
Besides the major abundances, mentioned above, explosive neon
burning supplies also substantial amounts
of $^{27}$Al, $^{29}$Si, $^{32}$S, $^{30}$Si, and $^{31}$P.
Explosive carbon burning contributes in addition
the nuclei $^{20}$Ne, $^{23}$Na, $^{24}$Mg, $^{25}$Mg, and $^{26}$Mg.
Many nuclei in the mass range 20$<$$A$$<$30 can be
reproduced in solar proportions. This was confirmed for realistic
stellar conditions by Morgan (1980). As photodisintegrations become
important in explosive Ne-burning, also heavier pre-existing nuclei in
such burning shells, from previous s- or r-processing (originating from
prior stellar evolution or earlier stellar generations), can undergo
e.g. $(\gamma,n)$ or $(\gamma,\alpha)$ reactions. These can produce
rare proton-rich stable isotopes of heavy elements. The relation to 
the so-called p-process is discussed e.g. in  Woosley \& Howard (1978),
Rayet, Prantzos, \& Arnould (1990), Howard, Meyer, \& Woosley (1991),
and Rayet et al. (1995).

\subsubsection{Explosive O-Burning}

Temperatures in excess of roughly $3.3\times 10^9$K lead to a 
quasi-equilibrium (QSE) in the lower QSE-cluster which extends over the range  
28$<$A$<$45 in mass number, while the path to heavier nuclei is blocked 
by small Q-values and reaction cross sections for reactions out of closed
shell nuclei with $Z$ or $N$=20 (see already Woosley, Arnett, \&
Clayton 1973 or Hix \& Thielemann 1997).
A full NSE with dominant abundances in the Fe-group cannot be attained.
The main burning products are $^{28}$Si,
$^{32}$S, $^{36}$Ar, $^{40}$Ca, $^{38}$Ar, and $^{34}$S, while
$^{33}$S, $^{39}$K, $^{35}$Cl, $^{42}$Ca, and $^{37}$Ar have
mass fractions of less than $10^{-2}$.
In zones with temperatures close to $4\times 10^9$K there 
exists some contamination
by the Fe-group nuclei $^{54}$Fe, $^{56}$Ni, $^{52}$Fe, $^{58}$Ni,
$^{55}$Co, and $^{57}$Ni. 

The abundance distribution within the QSE-cluster is determined
by alpha, neutron, and proton abundances. Because electron captures during
explosive processing are negligible, the original neutron excess stays 
unaltered and fixes the neutron to proton ratio. Under those conditions the 
resulting composition is dependent only 
on the alpha to neutron ratio at freeze-out. In an extensive study
Woosley, Arnett, and Clayton (1973) noted that
with a neutron excess $\eta$ of $2\times 10^{-3}$
the solar ratios of $^{39}$K/$^{35}$Cl  $^{40}$Ca/$^{36}$Ar, 
$^{36}$Ar/$^{32}$S, $^{37}$Cl/$^{35}$Cl, $^{38}$Ar/ $^{34}$S, 
$^{42}$Ca/$^{38}$Ar, $^{41}$K/$^{39}$K, and $^{37}$Cl/$^{33}$S are attained 
within a factor of 2 for freeze-out temperatures in the range 
$(3.1-3.9)\times 10^9$K.
This is the typical neutron excess resulting from solar CNO-abundances,
which are first transformed into $^{14}$N in H-burning and then into
$^{22}$Ne in He-burning via $^{14}$N($\alpha, 
\gamma$)$^{18}$F($\beta ^+$)$^{18}$O($\alpha ,\gamma$)$^{22}$Ne.
Similar results were obtained earlier by
Truran and Arnett (1970),
while for lower values of the neutron excess (as expected
for stars of lower metallicity) essentially only the alpha nuclei
$^{28}$Si, $^{32}$S, $^{36}$Ar, $^{40}$Ca are
produced in sufficient amounts (Truran and Arnett 1971). 
This affects element abundances and causes an odd-even staggering in Z.

\subsubsection{Explosive Si-Burning} 
\begin{figure}
\epsfig{file=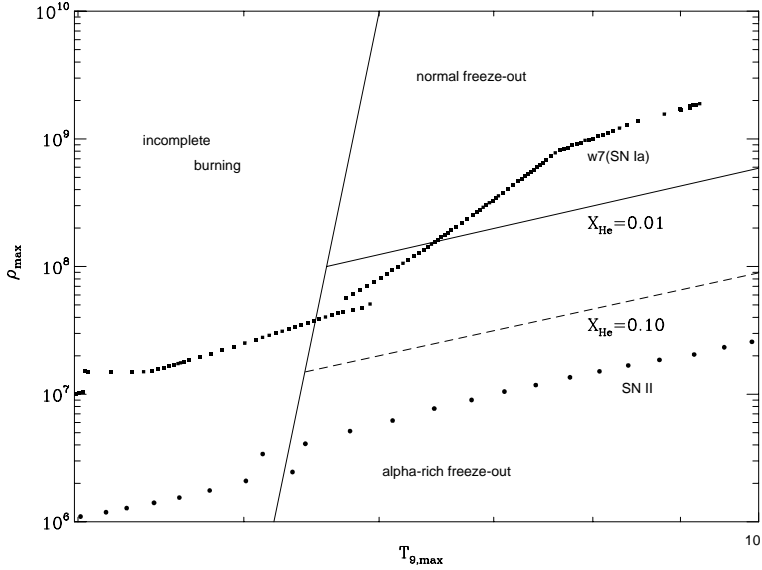,angle=90,width=5cm}
\vspace{2cm}
\caption{
Division of the $\rho_{max} - T_{max}$-plane for adiabatic expansions from
$\rho_{max}$ and $T_{max}$ with an adiabatic index of 4/3 and a hydrodynamic 
timescale equal to the free fall timescale. Conditions separate into
incomplete and complete Si-burning with normal and alpha-rich freeze-out.
Contour lines of constant $^4$He mass fractions in complete burning are
given for levels of 1 and 10\%. They coincide with lines of constant radiation
entropy per gram of matter. For comparision also the maximum
$\rho-T$-conditions of individual mass zones in type Ia and type II supernovae
are indicated.
\label{incompletesi}} 
\end{figure}

Zones which experience temperatures in excess of 4.0--5.0$\times 10^9$K
undergo explosive Si-burning. For $T$$>$5$\times 10^9$K essentially all 
Coulomb barriers can be overcome and a nuclear statistical equilibrium is
established. Such temperatures lead to complete Si-exhaustion and produce 
Fe-group nuclei. The doubly-magic nucleus $^{56}$Ni, with the largest
binding energy per nucleon for $N$=$Z$, is formed with a dominant abundance
in the Fe-group in case $Y_e$ is larger than 0.49.
Explosive Si-burning can be devided into three different regimes:
(i) incomplete Si-burning and complete Si-burning with
either (ii) a normal or (iii) an alpha-rich freeze-out.
Which of the three regimes is encountered depends on the peak temperatures
and densities attained during the passage of supernova shock front (see Fig.~20
in Woosley, Arnett, and Clayton 1973, and for applications to supernova
calculations Fig.~4 in Thielemann et al.~1996b and Fig.~5 in
Thielemann, Hashimoto, \& Nomoto 1990 -- combined here as 
Fig.~\ref{incompletesi}). One recognizes that the mass zones of SNe Ia and 
SNe II experience different regions of complete Si-burning. 

\begin{figure}
\epsfig{file=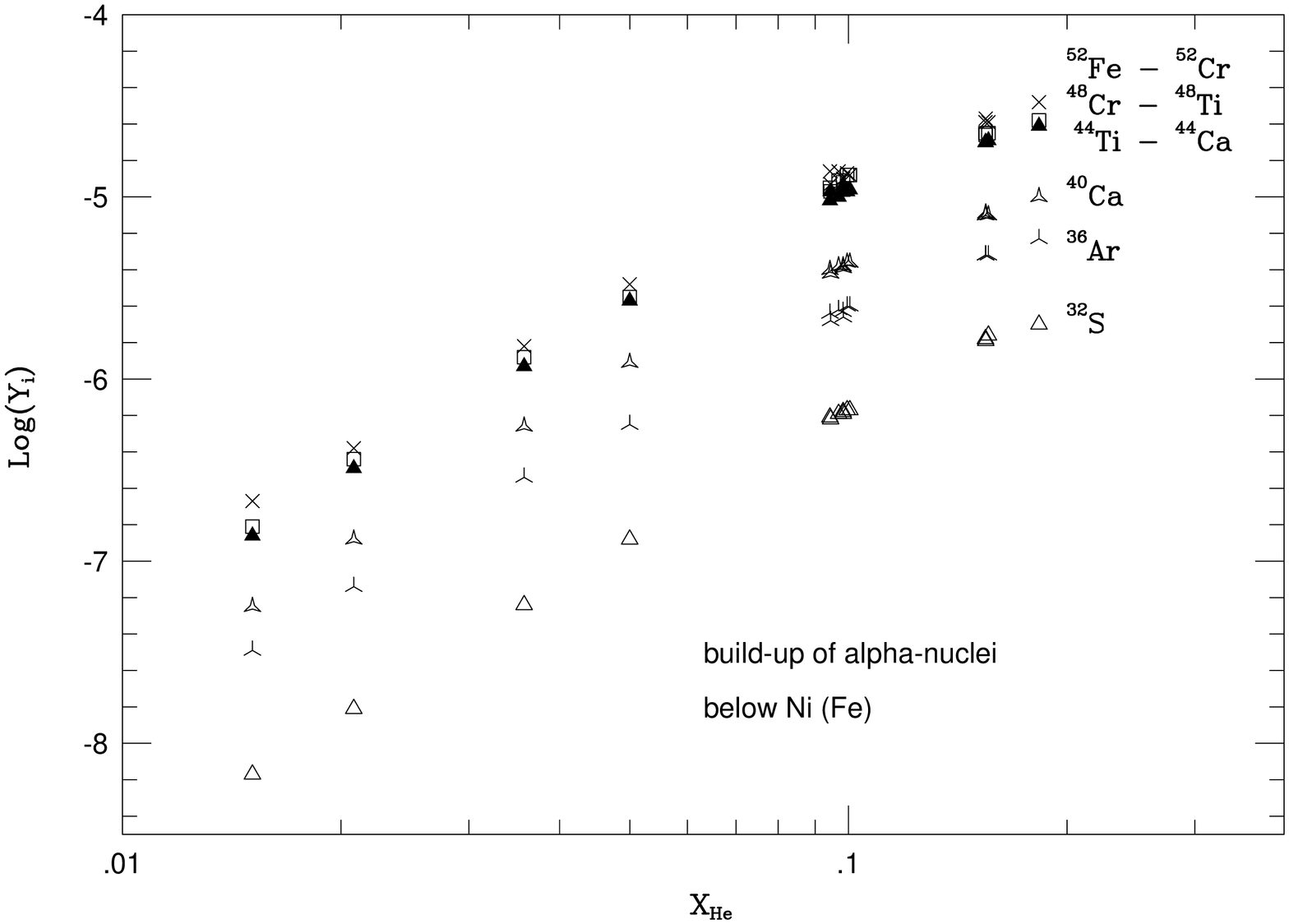,angle=90,width=12cm}
\caption{Display of the composition up to Cr and from
Mn to Ni (after decay) as a function of remaining alpha mass-fraction
$X_{He}$ from an alpha-rich freeze-out. Lighter nuclei, being produced
by alpha-captures from a remaining alpha reservoir, have larger abundances
for more pronounced alpha-rich freeze-outs. Nuclei beyond Fe and Ni
behave similarly, because of shifts in the dominant Fe-group nuclei
caused by alpha-captures. Therefore, the dominant Fe-group nuclei like Fe and
Ni show the opposite effect.
\label{alpharich}}
\vspace{-2.5cm}
\epsfig{file=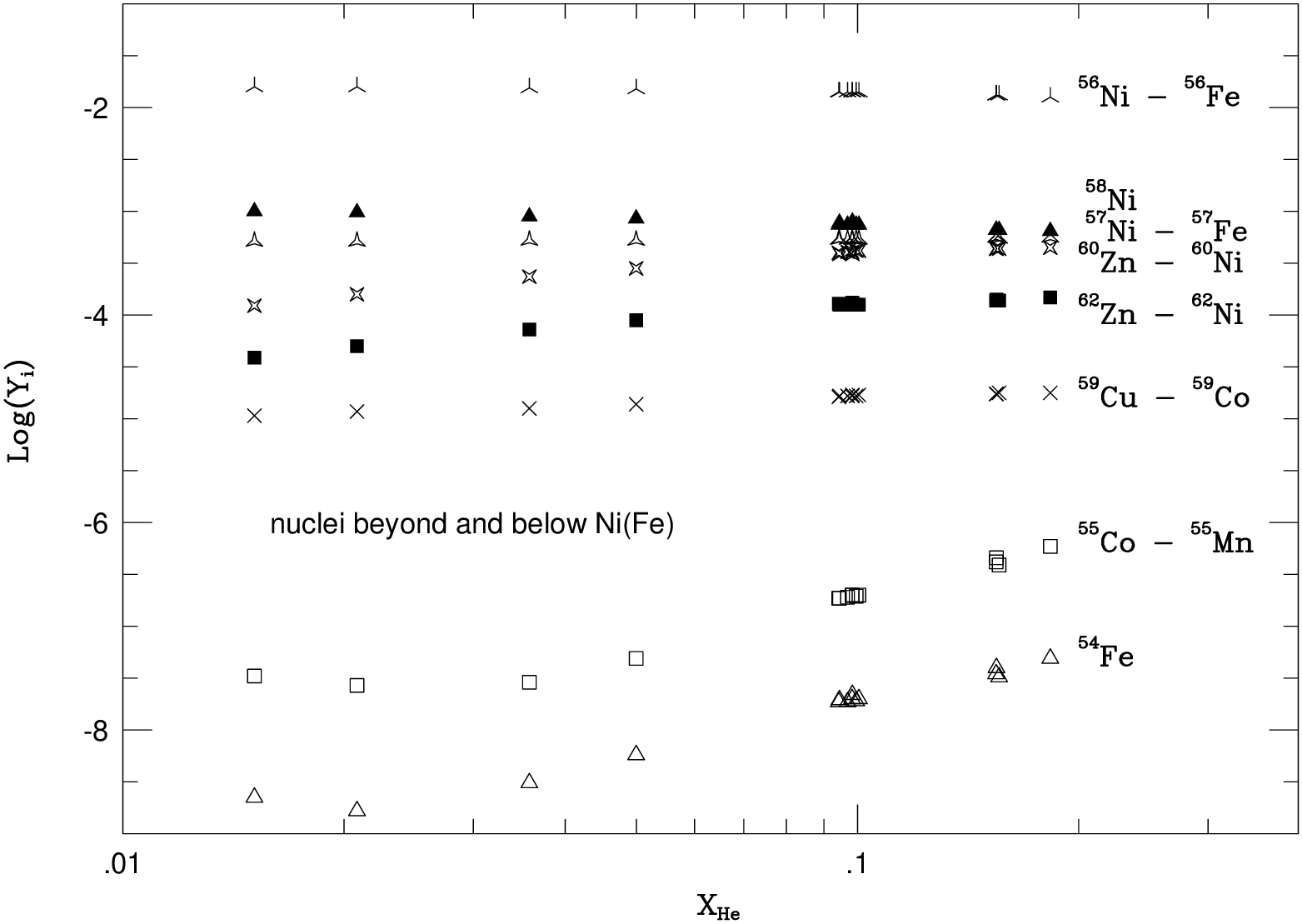,angle=0,width=9.3cm}
\end{figure}

At high temperatures in complete Si-burning or also during a normal freeze-out,
the abundances are in a full NSE and given by Eq.(\ref{nseeq}). 
An alpha-rich freeze-out is caused by the inability of the triple-alpha
reaction $^4$He(2$\alpha,\gamma)^{12}$C, transforming $^4$He into $^{12}$C,
and the $^4$He($\alpha n,\gamma)^9$Be reaction, to keep light nuclei like $n$, 
$p$, and $^4$He, and intermediate mass nuclei beyond $A$=12 in an NSE during 
declining temperatures, when the densities are small. The latter enter 
quadratically for these rates, causing during the fast expansion and cooling in 
explosive events a large alpha abundance after charged particle freeze-out,
which shifts the QSE groups to heavier nuclei, tranforming e.g. $^{56}$Ni, 
$^{57}$Ni, and $^{58}$Ni into $^{60}$Zn, $^{61}$Zn, and $^{62}$Zn.
This also leads to a slow supply of carbon nuclei still during freeze-out, 
leaving traces of alpha nuclei, $^{32}$S, $^{36}$Ar, $^{40}$Ca, $^{44}$Ti, 
$^{48}$Cr, and $^{52}$Fe, which did not fully make their way up to $^{56}$Ni.
Figs.~\ref{alpharich}ab show this effect, typical for SNe II, 
as a function of remaining alpha-particle mass fraction after
freeze-out. It is clearly seen that the major NSE nuclei
$^{56}$Ni, $^{57}$Ni, and $^{58}$Ni get depleted when the remaining
alpha fraction increases, while all other species mentioned above
increase.  

Incomplete Si-burning is characterized 
by peak temperatures of $4-5\times 10^9$K. Temperatures are not high enough
for an efficient bridging of the bottle neck above the proton magic number
$Z$=20 by nuclear reactions. Besides the dominant fuel
nuclei $^{28}$Si and $^{32}$S we find the alpha-nuclei $^{36}$Ar and
$^{40}$Ca being most abundant. 
Partial leakage through the bottle neck above $Z$=20 produces
$^{56}$Ni and $^{54}$Fe as dominant abundances in the Fe-group.
Smaller amounts of $^{52}$Fe, $^{58}$Ni,
$^{55}$Co, and $^{57}$Ni are encountered. 
All explosive burning phases discussed above will be applied in more
detail to SNe II nucleosynthesis in section 4.

\subsubsection{The r-Process}  

The operation of an r-process is characterized by the fact that
10 to 100 neutrons per seed nucleus (in the Fe-peak or somewhat beyond) have 
to be available to form all heavier r-process nuclei by neutron capture. 
For a composition of Fe-group nuclei and free neutrons that
translates into a neutron excess of $\eta=0.4-0.7$ or $Y_e$=0.15-0.3. Such a 
high neutron excess can only by obtained through capture of energetic
electrons 
(on protons or nuclei) which have to overcome large negative Q-values.
This can be achieved by degenerate electrons with large Fermi
energies and requires a compression to densities of
$10^{11}-10^{12}$g cm$^{-3}$, with a beta equilibrium between electron 
captures and $\beta^-$-decays (Cameron 1989) as found in neutron star
matter (see also Meyer 1989). 

Another option is an extremely alpha-rich freeze-out in complete
Si-burning with moderate neutron excesses $\eta$ and $Y_e$'s (0.16 or 0.42,
respectively). After the freeze-out of charged particle reactions
in matter which expands from high temperatures but relatively low
densities, 70, 80, 90 or 95\% of all matter can be locked into $^4$He with
$N$=$Z$. Figure \ref{alpharich} showed the onset of such an extremely alpha-rich
freeze-out by indicating contour lines for He mass fractions
of 1 and 10\%. These contour lines correspond to $T_9^3/\rho$=const,
which is proportional to the entropy per gram of matter of a radiation
dominated gas. Thus, the radiation entropy per gram of baryons can be
used as a measure of the ratio between the remaining He mass-fraction and
heavy nuclei. Similarly, the ratio of neutrons to Fe-group (or heavier) nuclei 
(i.e. the neutron to seed ratio) is a function of entropy and permits for
high entropies, with large remaining He and neutron abundances and small
heavy seed abundances, neutron captures which proceed 
to form the heaviest r-process nuclei (Woosley \& Hoffman 1992, Meyer et al.
1992, Takahashi et al. 1994, Woosley et al. 1994b, Hoffman et al. 1996, 1997). 

A different situation surfaces for maximum temperatures below
freeze-out conditions for charged particle reactions with Fe-group
nuclei. Then reactions among light nuclei which release neutrons,
like $(\alpha,n)$ reactions on $^{13}$C and $^{22}$Ne, can sustain a
neutron flux. The constraint of
having 10-100 neutrons per heavy nucleus, in order to attain r-process
conditions, can only be met by small abundances of Fe-group nuclei.
Such conditions were expected when a shock front passes the He-burning
shell and enhances the $^{22}$Ne($\alpha,n)$ reaction by orders of
magnitude.
However, Blake et al. (1981) and Cowan, Cameron, \& Truran (1983) could
show that this neutron source is not strong enough for an r-process in
realistic stellar models (see also subsection 3.2.1).
Recent research, based on additional neutron release via 
inelastic neutrino scattering (Epstein, Colgate, and Haxton 1988), can 
also not produce neutron densities which are required for such
a process to operate (see also Woosley et al. 1990 and Meyer 1995).

r-process calculations independent of a specific astrophysical site, and
just based on the goal to find the required neutron number densities and
temperatures which can reproduce the solar abundance pattern of heavy
elements, have been performed for a number of years. The latest ones are
e.g. Kratz et al. (1988), Thielemann et al. (1993), Kratz et al. (1993), 
Thielemann et al. (1994a), Chen et al. (1995), Bouquelle
et al. (1996), Pfeiffer, Kratz, \& Thielemann (1997), Kratz, Pfeiffer,
\& Thielemann (1997) and Freiburghaus et al. (1997b). They, together with 
applications to the astrophysical sites listed 
above, will be discussed in section 5.

\subsection{Nucleosynthesis in Supernovae}

In the following section 4 we will apply these explosive burning processes
to nucleosynthesis calculations in supernova explosions from massive
stars (SNe II) for nuclei with A$<$70. The discussion of the explosive 
production of heavier nuclei in supernovae will be given in section 5.
There exist many original and review articles about the mechanisms of
SNe II (e.g. Bruenn 1989ab, Cooperstein \& Baron 1990,
Wilson \& Mayle 1988, Mayle \& Wilson 1990, Bethe 1990, Bruenn \& Haxton 1991;
Wilson \& Mayle 1993, Herant et al. 1994, Janka \& M\"uller 1995,1996, 
Burrows 1996; Mezzacappa et al. 1997), so that we do not intend to repeat 
such a discussion here. We rather want to concentrate on the accompanying 
nucleosynthesis processes. 

One of the major free parameters in stellar evolution, and thus for the
pre-supernova models, is the still uncertain $^{12}$C($\alpha ,\gamma)^{16}$O
reaction (see Filippone, Humblet, \& Langanke 1989, Caughlan et al.
1985, Caughlan and Fowler 1988, Barker \& Kajino 1991,
Buchmann et al. 1993, Zhao et al. 1993ab, Azuma et al. 1994, 
Langanke and Barnes 1996, Buchmann et al. 1996,1997).
The permitted uncertainty range still spans almost over a factor of 3.
However, also the treatment of convection in stellar evolution is not a
settled one, especially the issue of overshooting and semiconvection. This has 
an influence on the possible growth of the He-burning core, which causes 
mixing in of fresh He at higher
temperatures, and consequently also enhances the O/C ratio.
Stellar evolution calculations by Langer and Henkel (1995) show that the total 
amount of $^{16}$O can also vary by almost a factor of 3, for the extreme
choices of the semi-convection parameter. 

Thus, only the combination of these two uncertain parameters can be determined
by comparison with abundance observations from supernova explosions.
The calculations, presented in this review, are based on stellar
models which made use of the Schwarzschild criterion of convection 
(Nomoto \& Hashimoto 1988; Hashimoto et al. 1995) and
employed the $^{12}$C($\alpha,\gamma)^{16}$O-rate by Caughlan et al. (1985), 
which is one choice within the permitted uncertainty window.

\section{Type II Supernova Explosions}

All stars with main sequence masses M$>$8M$_\odot$ 
(e.g. Nomoto \& Hashimoto 1988, Hashimoto, Iwamoto \& Nomoto 1993, Weaver \&
Woosley 1993) produce a collapsing core after the end of their hydrostatic 
evolution, which proceeds to nuclear densities (for a review see e.g. Bethe 
1990). The total energy released, 2-3$\times 10^{53}$erg, equals the
gravitational binding energy of a neutron star. Because neutrinos are the
particles with the longest mean free path, they are able to carry away that
energy in the fastest fashion. This was proven by the neutrino emission of 
supernova 1987A, detected in the Kamiokande, IMB and Baksan experiments 
(see Burrows 1990 for an overview).

The most promising mechanism for supernova explosions is based on 
neutrino heating beyond the hot proto-neutron star via the dominant processes 
$\nu_e + n \rightarrow p+e^-$ and $\bar\nu_e+p\rightarrow n+e^+$ with a 
(hopefully) about 1\% efficiency in energy deposition (see also M. Guidry, 
this volume). The neutrino heating efficiency depends on the neutrino 
luminosity, which in turn is affected by neutrino opacities (e.g. Bruenn 1985, 
Sawyer 1989, Schinder 1990, Horowitz \& Wehrberger 1992, Mezzacappa \& Bruenn 
1993, Keil \& Janka 1995, Reddy \& Prakash 1997, Reddy, Prakash, \& Lattimer 
1997). The explosion via neutrino heating is delayed after core collapse
for a timescale of seconds or less. The exact delay time $t_{de}$
depends on the question whether neutrinos diffuse out from the core ($>$0.5s),
weak convection occurs due to composition gradients,
or convective turnover due to entropy gradients shortens this escape time
substantially (e.g. Burrows \& Fryxell 1992, Janka \& M\"uller 1993,  
Wilson \& Mayle 1993, Herant et al. 1994, Bruenn, Mezzacappa, \& Dineva 1995,
Janka \& M\"uller 1995,1996, Burrows 1996; Mezzacappa et al. 1997).
The behavior of $t_{de}$ as a function of stellar mass is still an open         
question and quantitative results of self-consistent calculations 
should still be taken with care, suggesting instead to
make use of the fact that typical kinetic energies of             
$10^{51}$ erg are observed and light curve as well as explosive nucleosynthesis 
calculations can be performed by introducing a shock of 
appropriate energy in the pre-collapse stellar model (see e.g. Woosley 
\& Weaver 1986, Shigeyama, Nomoto \& Hashimoto 1988, Thielemann, Hashimoto,
\& Nomoto 1990, Aufderheide, Baron, \& Thielemann 1991, Weaver and Woosley
1993, Woosley \& Weaver 1995, Thielemann, Nomoto, \& Hashimoto 1996,
Nomoto et al. 1997). 
Due to these remaining open questions, 
present explosive nucleosynthesis calculations for SNe II are still 
based on such induced supernova explosions by either depositing thermal energy
or invoking a piston with a given kinetik energy of the order $10^{51}$ erg,
in order to process and eject matter outside the collapsed Fe-core of a 
massive star. 

These are not self-consistent calculations, which would also precisely
determine a mass cut between the 
central neutron star and the ejected envelope. Although self-consistent
calculations show promising results in recent years, on the one hand one expects
changes from 2D to more realistic 3D calculations, on the other hand
issues like the mass cut are not consistently solved yet, and some models
would eject very unwanted nucleosythesis products. Induced calculations,
with the constraint of requiring ejected $^{56}$Ni-masses from the innermost
explosive Si-burning layers in agreement with supernova light curves,
being powered by the decay chain $^{56}$Ni-$^{56}$Co-$^{56}$Fe,
are preferable at this point and can also serve as guidance to the solution of 
the whole supernova problem. 
Such mass cuts, based on $^{56}$Ni in the ejecta,
are always the "final" cuts, not necessarily the position of the high entropy
bubble where neutrino heating causes the explosion. Massive stars will have
some fallback, caused by reverse shocks reflected at density jumps in the
outer layers. Recent observations of massive type II supernovae with
very small amounts of $^{56}$Ni are an indication for just this effect
(Schmidt 1997, Turatto et al. 1997, Sollerman et al. 1998).
Thus, when we will use the expression mass cut in the following, it will
always relate to the final cut after fallback.

The composition of the innermost ejected layers is crucial and
reflects aspects of the total energy in the shock and the temperatures 
attained due to it (responsable for $^{56}$Ni), the neutronization of matter
in form of $Y_e$, affecting the Fe-group composition in general and especially
the $^{57}$Ni/$^{56}$Ni ratio, and finally the entropy of the material
which determines the degree of the alpha-rich freeze-out and with it the amount
of some intermediate-mass alpha-elements like radioactive $^{44}$Ti. 
Comparison with abundances from specific supernova observations or supernova 
remnants can teach a lot about these details and the supernova
mechanism as a function of progenitor mass. The amount of detected $^{16}$O 
and $^{12}$C or products                
from carbon and explosive oxygen burning can constrain our knowledge of the     
{\it effective} $^{12}$C($\alpha,\gamma)^{16}$O                                 
rate in He-burning. The $^{57}$Ni/$^{56}$Ni ratio                  
can give constraints on $Y_e$ in the innermost ejected zones. 
This helps to estimate the necessary delay time between collapse and the
neutrino-driven explosion. Provided that the stellar  
pre-collapse models are reliable, this allows additional insight into the exact 
working of the supernova explosion mechanism.          

All these aspects can be explored when
being guided by comparison to observations (e.g.
SN1987A, a 20M$_\odot$ during the main sequence stage -- see e.g. Arnett et 
al. 1989, McCray 1993, Fransson \& Kozma 1993, Suntzeff et al. 1992, 1997,
Kozma \& Fransson 1997; SN 1993J, a 14$\pm$1M$_\odot$ star during main sequence
-- see e.g. Nomoto et al. 1993, Shigeyama et al. 1994, Woosley et al. 1994a,
Houck \& Fransson 1996;   
type Ib and Ic supernova light curves like SN 1994I, which due to the lack of 
a large H-envelope and their early X-ray and gamma-ray losses are steeper 
than those of SNe II, but are also core collapse events -- see e.g. Shigeyama
et al. 1990, Nomoto et al. 1994, Iwamoto et al. 1994;                      
the $^{57}$Ni/$^{56}$Ni ratio deduced from                  
$\gamma$-rays of the $^{56,57}$Co decay or from spectral features changing    
during the decay time -- see e.g. Clayton et al. 1992, Kurfess et al. 1992,
Kumagai et al. 1993, Fransson \& Kozma 1993, Varani et al. (1990); or 
supernova remnants like G292.0+1.8, N132D, CAS A -- Hughes and Singh 1994, 
Blair et al. 1994, Iyudin et al. (1994), Dupraz et al. (1997), Hartmann et al. 
(1997); and comparison with abundances in low metallicity 
stars, which reflect the average SNe II composition (Wheeler, Sneden, \& Truran
1989, Lambert 1989, Pagel 1991, Zhao \& Magain 1990, Gratton \& Sneden 1991,
Edvardsson et al. 1993, Nissen et al. 1994, McWilliam et al. 1995,
Schuster et al. 1996, Ryan, Norris, \& Beers 1996, Norris, Ryan, \& Beers 1996,
Barbuy er al. 1997, Beers, Ryan, \& Norris 1997, McWilliam 1997).

We concentrate here on the composition of 
the ejecta from such core collapse supernovae as an extension to            
earlier work (Hashimoto, Nomoto \& Shigeyama 1989;       
Thielemann, Hashimoto \& Nomoto 1990; Thielemann, Nomoto \& Hashimoto 1993, 
1994, 1996; Hashimoto et al. 1993, 1995; and Nomoto et al. 1997).           
                                                                                
\subsection{Basic Nucleosynthesis Features}                            
                                                                                
\noindent 
The calculations were performed by depositing a total thermal                   
energy of the order $E=10^{51}$erg + the gravitational binding energy of        
the ejected envelope into several mass zones of the stellar Fe-core.            
A first overview of results from the explosion calculations          
can be seen in Table 3 of Thielemann et al. (1996a) for element abundances
and Table 1 of Nomoto et al. (1997) for isotopic abundances in the
supernova ejecta as a function of progenitor star mass.
They can be characterized by the following behavior:          
the amount of ejected mass from the unaltered (essentially   
only hydrostatically processed) C-core and from explosive Ne/C-burning          
(C, O, Ne, Mg)                                                                  
varies strongly over the progenitor mass range, while the amount of mass        
from explosive O- and Si-burning (S, Ar, and Ca) is almost the                  
same for all massive stars. Si has some contribution from hydrostatic           
burning and varies by a factor of 2-3.                                          
The amount of Fe-group nuclei ejected                                           
depends directly on the explosion mechanism. 
The values listed for the 20M$_\odot$ star have been
chosen to reproduce the 0.07M$_\odot$ of $^{56}$Ni deduced from light curve     
observations of SN 1987A. The        
choice for the other progenitor masses is also based on supernova light
curve observations, but their uncertain nature should be underlined 
and a clearer picture is only emerging now with the observation of
varying amounts of $^{56}$Ni for varying progenitor star masses
(see Blanton, Schmidt, \& Kirshner 1995, Schmidt 1997, Turatto et al. 1997, 
Sollerman, Cumming, \& Lundquist 1998).                            

Thus, we have essentially          
three types of elements, which test different aspects of supernovae,            
when comparing with individual observations. (i) The first set (C, O, Ne,    
Mg) tests the stellar progenitor models, (ii) the second (Si, S, Ar, Ca) the   
progenitor models and the explosion energy in the shock wave, while (iii) the 
Fe-group (beyond Ti) probes clearly in addition the actual supernova            
mechanism. Only when all three aspects of the predicted abundance yields        
can be verified with individual observational checks, it will be                
reasonably secure to utilize these results in chemical evolution                
calculations of galaxies (see e.g. Tsujimoto et al. 1995; Timmes et al.
1995; Pagel \& Tautvaisiene 1995, 1997; Tsujimoto et al. 1997).                                    
In general we should keep in mind, that as long as the explosion mechanism is 
not completely and quantitatively understood yet, one has to assume a position 
of the mass cut which causes (not predicts!) a specific amount of $^{56}$Ni 
ejecta. Dependent on that position, which     
is a function of the delay time between collapse and final explosion,           
the ejected mass zones will have a different neutron excess
or $Y_e$=$<Z/A>$ of the nuclear composition, determining the ratio 
$^{57}$Ni/$^{57}$Ni. The nature and amount of the energy deposition affects
the entropy in the innermost ejected layers, and with it the degree of the
alpha-rich freeze-out and amount of $^{44}$Ti ejecta. We will discuss    
this in more detail in the following subsections.                               
                                                                                
\subsection{Ni(Fe)-Ejecta and the Mass Cut}                   

\begin{figure}
\vspace{-2cm}
\epsfig{file=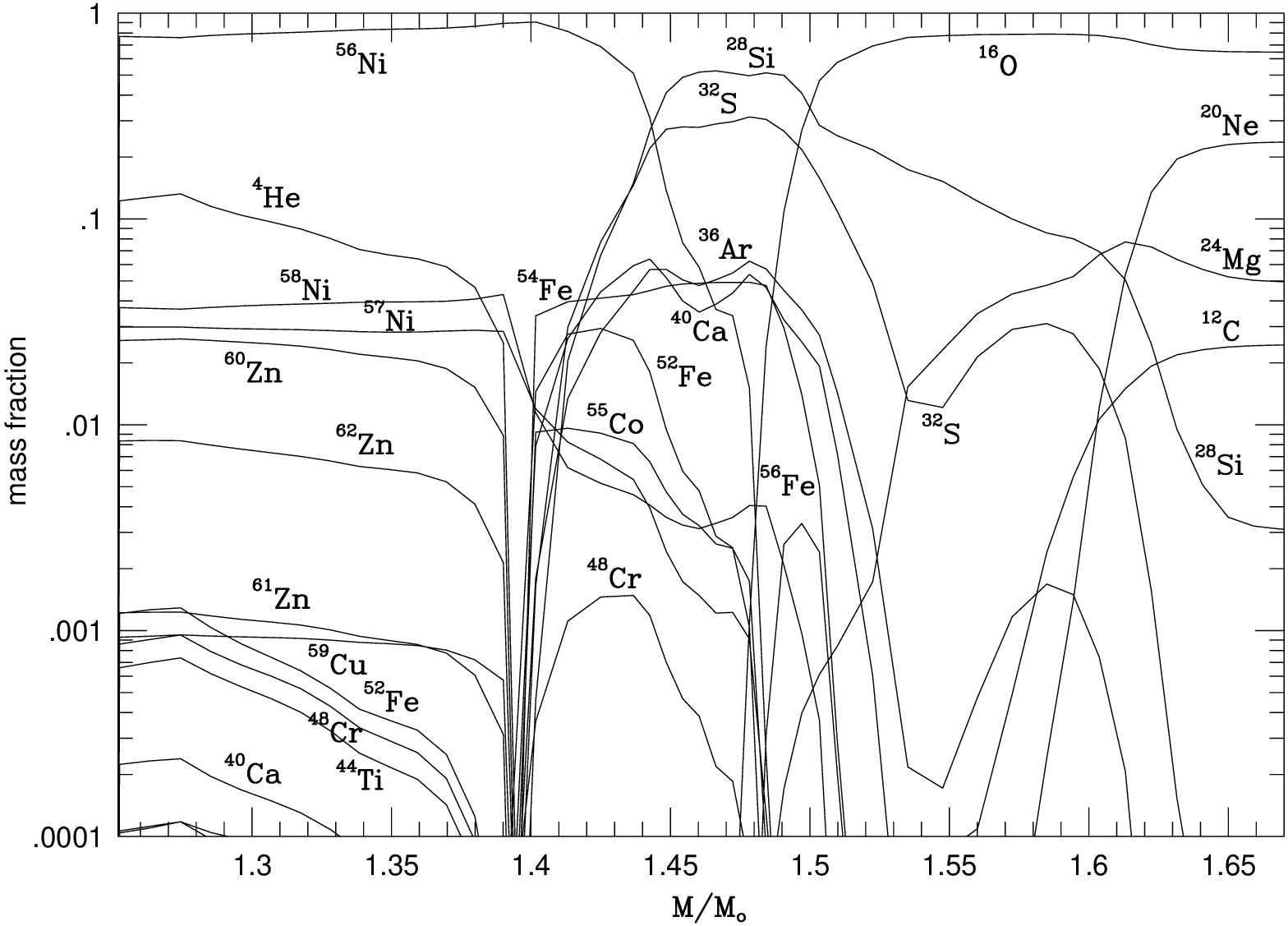,width=9cm}
\caption{Isotopic composition of the ejecta for a core collapse          
supernova from a 13M$_\odot$ star (3.3M$_\odot$   
He-core). Only the dominant abundances  
of intermediate mass nuclei are plotted, while the Fe-group composition         
is presented in full detail.            
The exact mass cut in $M(r)$ between neutron star and ejecta                
depends on the details of the delayed explosion mechanism.                 
Figures 10a and 10b show how strongly a $Y_e$-change 
can affect the resulting composition. Figure 10a makes use of a constant  
$Y_e$=0.4989 in the inner ejcta, Figure 10b makes use of the original $Y_e$,    
resulting from the pre-collapse burning phases, which drops to 0.4915
at the position for matter resulting from core O-burning, which experienced
high densities and electron captures.
\label{expnuc}}
\vspace{-2cm}
\epsfig{file=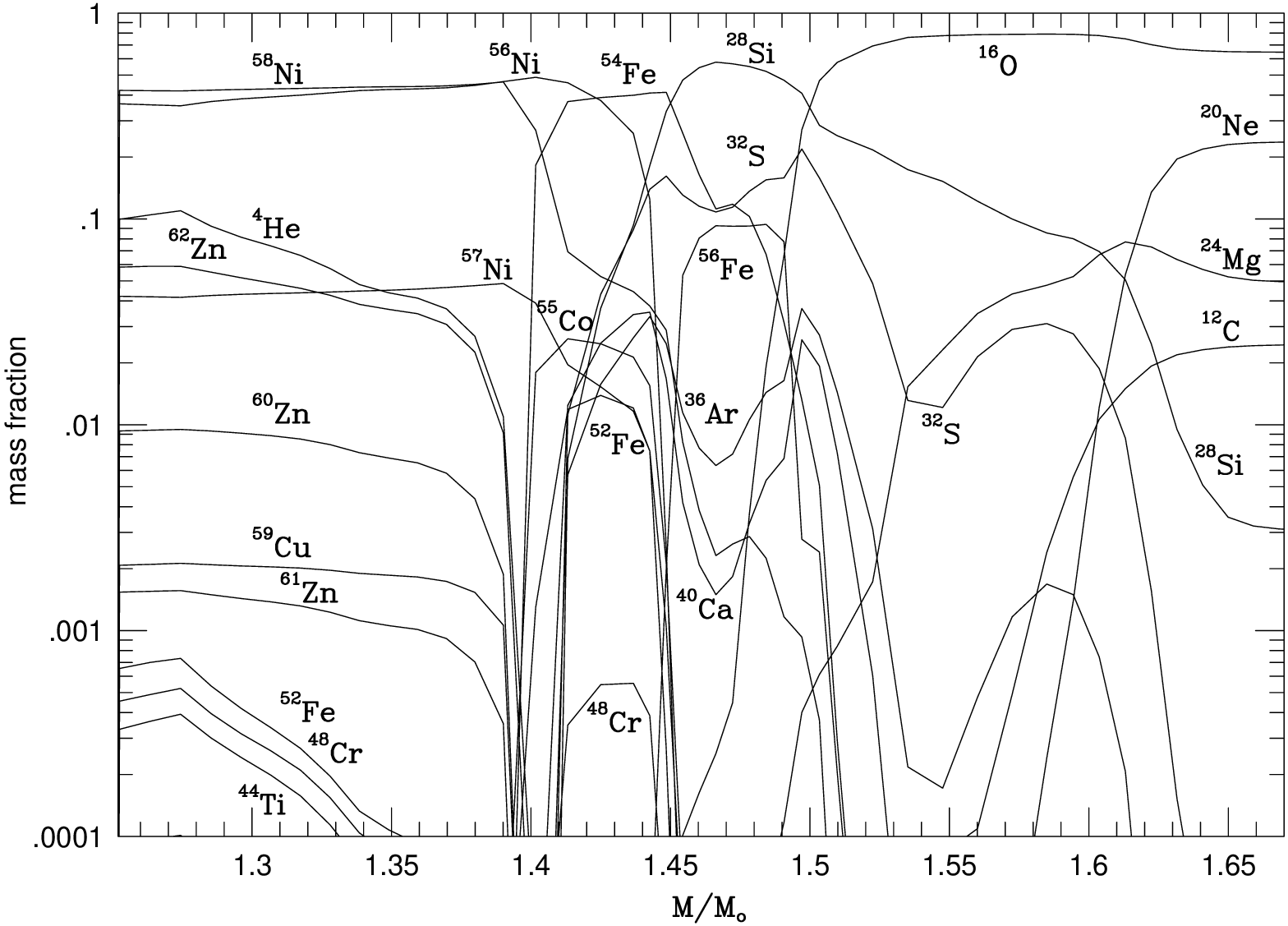,width=9cm}
\end{figure}

Figs.~\ref{expnuc}ab (both presenting a 13M$_\odot$ star) make clear how   
strongly a $Y_e$ change can                                                     
affect the resulting composition. Fig.~\ref{expnuc}a makes use of a constant 
$Y_e$=0.4989 in the inner ejcta, experiencing incomplete and complete           
Si-burning. Figure \ref{expnuc}b makes use of the original $Y_e$, resulting from
the pre-collapse burning phases. Here $Y_e$ drops to 0.4915 for mass zones    
below $M(r)$=1.5M$_\odot$. Huge changes in the Fe-group composition can         
be noticed. The             
change in $Y_e$ from 0.4989 to 0.4915 causes a tremendous change in the         
isotopic composition of the Fe-group for the affected mass regions              
($<$1.5M$_\odot$). In the latter case the abundances of $^{58}$Ni and           
$^{56}$Ni become comparable. All neutron-rich isotopes increase                 
($^{57}$Ni, $^{58}$Ni, $^{59}$Cu, $^{61}$Zn, and $^{62}$Zn), the even-mass 
isotopes ($^{58}$Ni and $^{62}$Zn) show the strongest effect.              
In Fig.~\ref{expnuc} one can also recognize        
the increase of $^{40}$Ca, $^{44}$Ti, $^{48}$Cr, and $^{52}$Fe with an          
increasing remaining He mass fraction. These are direct consequences            
of a so-called alpha-rich freeze-out with increasing entropy.               
           
While these calculations were performed by depositing energy at a specific     
radius inside the Fe-core and letting the shock wave propagate outward,         
this should involve the outer structure of the star after collapse and a
the time $t_{de}$, when the successful shock wave is initiated.     
Instead they were taken at the onset of core collapse, which would 
corresponds to a prompt explosion without delay. In case of a delayed        
explosion, accretion onto the proto-neutron star will occur until               
finally after a delay period $t_{de}$ a shock wave is formed, leading to the    
ejection of the outer layers. Aufderheide et al. (1991) performed a calculation
with a model at $t_{de}$=0.29s after core collapse for a 20M$_\odot$ star, when
the prompt shock had failed, and found an accretion caused increase of the 
mass cut by roughly $\Delta M_{acc}$=0.02M$_\odot$. A delayed explosion could 
set in after a delay of up to          
1s, with the exact time being somewhat uncertain and dependent on the details   
of neutrino transport (Wilson \& Mayle 1993, Herant et al. 1994, Bruenn,
Mezzacapp, \& Dineva 1995, Janka \& M\"uller 1996, Burrows 1996).                                                          
                                                                                
The outer boundary of explosive Si-burning with complete Si-exhaustion is given 
by $T$=5$\times$10$^9$K and is also the 
outer boundary of $^{56}$Ni production. From pure energetics it can be 
shown that this corresponds approximately to a radius $r_5$=3700 km         
for $E_{SN}$$\approx$10$^{51}$ erg, independent
of the progenitor models (Woosley 1988, Thielemann, Hashimoto, \& Nomoto 1990). 
Therefore, the mass cut would be at                                      
                                                                                
\begin{equation}
M_{cut}=M(r_5)-M_{ej}(^{56}{\rm Ni}).                              
\end{equation}
In case of a delayed explosion, we have to ask the question from which          
radius $r_{0,5}(t=0)$ matter fell in, which is located at radius                
$r_5(t$=$t_{de})$=3700km when the                                               
shock wave emerges at time $t_{de}$. This effect of accretion as a function
of delay time $t_{de}$ has been studied in detail (Thielemann et al.
1996a). Here we want to present only the quantitative results. 
\begin {figure}
\vspace{-2cm}
\epsfig{file=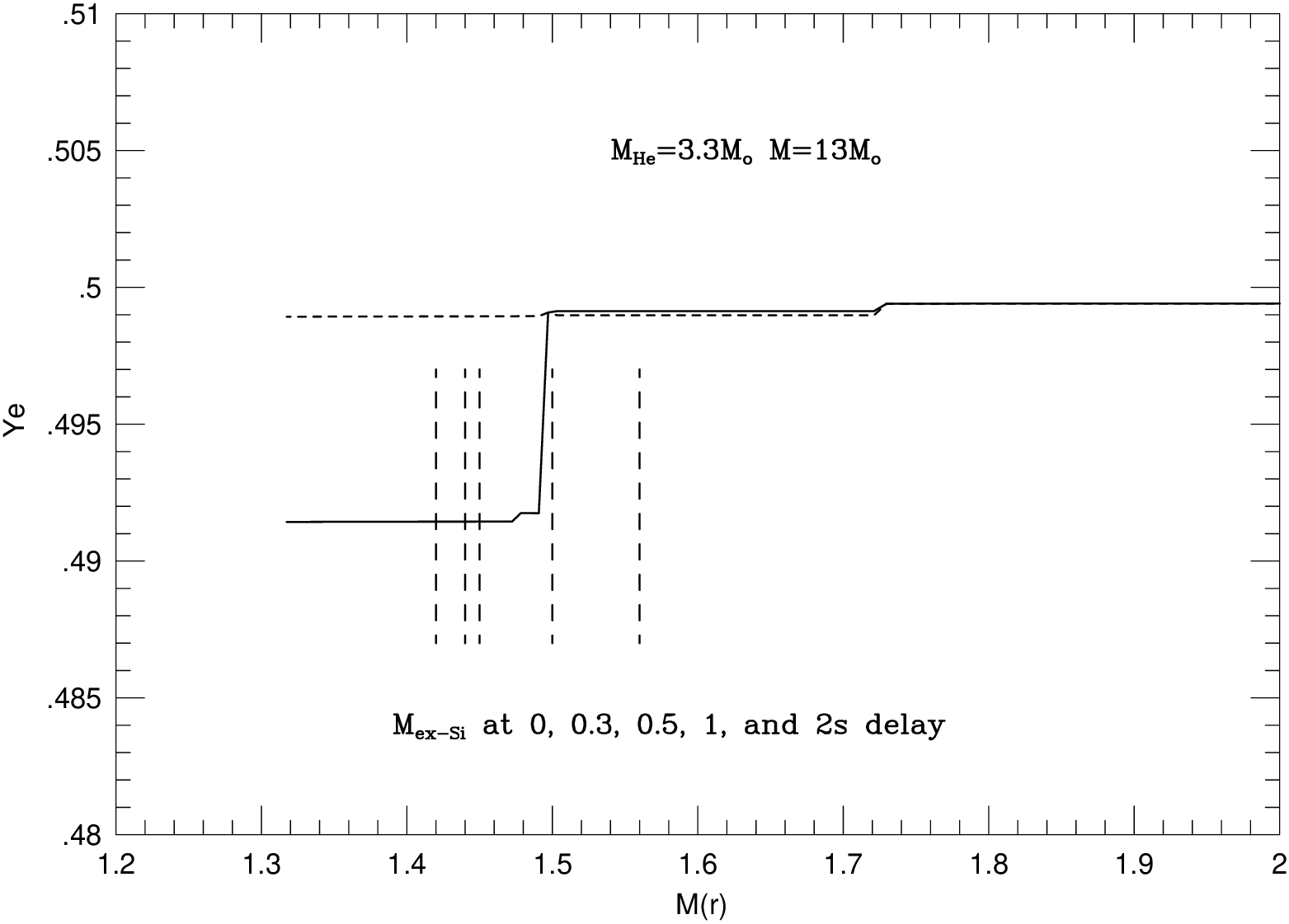,width=9cm}
\caption{Figures 11ab present the $Y_e$-distributions of a 13 and 20M$_\odot$
star and the position of the outer boundary of explosive                   
Si-burning with complete Si-exhaustion, $M_{ex-Si}$, as a function of           
the delay/accretion period $t_{de}$.                                            
For each star delay times of 0, 0.3, 0.5, 1, and 2s are considered, resulting   
in $r_{0,5}$=3700, 4042, 4412, 5410, and 7348km.          
$^{56}$Ni is produced inside this boundary $r_{0,5}$ as the dominant nucleus.   
For a given amount of Ni-ejecta, mass cuts would have to be positioned          
at $M_{cut}$=$M(r_{ex-Si})-M_{ej}(^{56}$Ni)=$M(r_{0,5}(0))-M_{ej}(^{56}$Ni). The
delay times $t_{de}$ and the required $M_{ej}(^{56}$Ni) determine $Y_e$ in the
ejected material (solid=original, dashed=experienced for sufficiently large
$t_{de}$, when low $Y_e$-matter is accreted onto the neutron star.
The steep drop in $Y_e$ corresponds to the edge of core O-burning.
\label{accneut}}          
\vspace{-2cm}
\epsfig{file=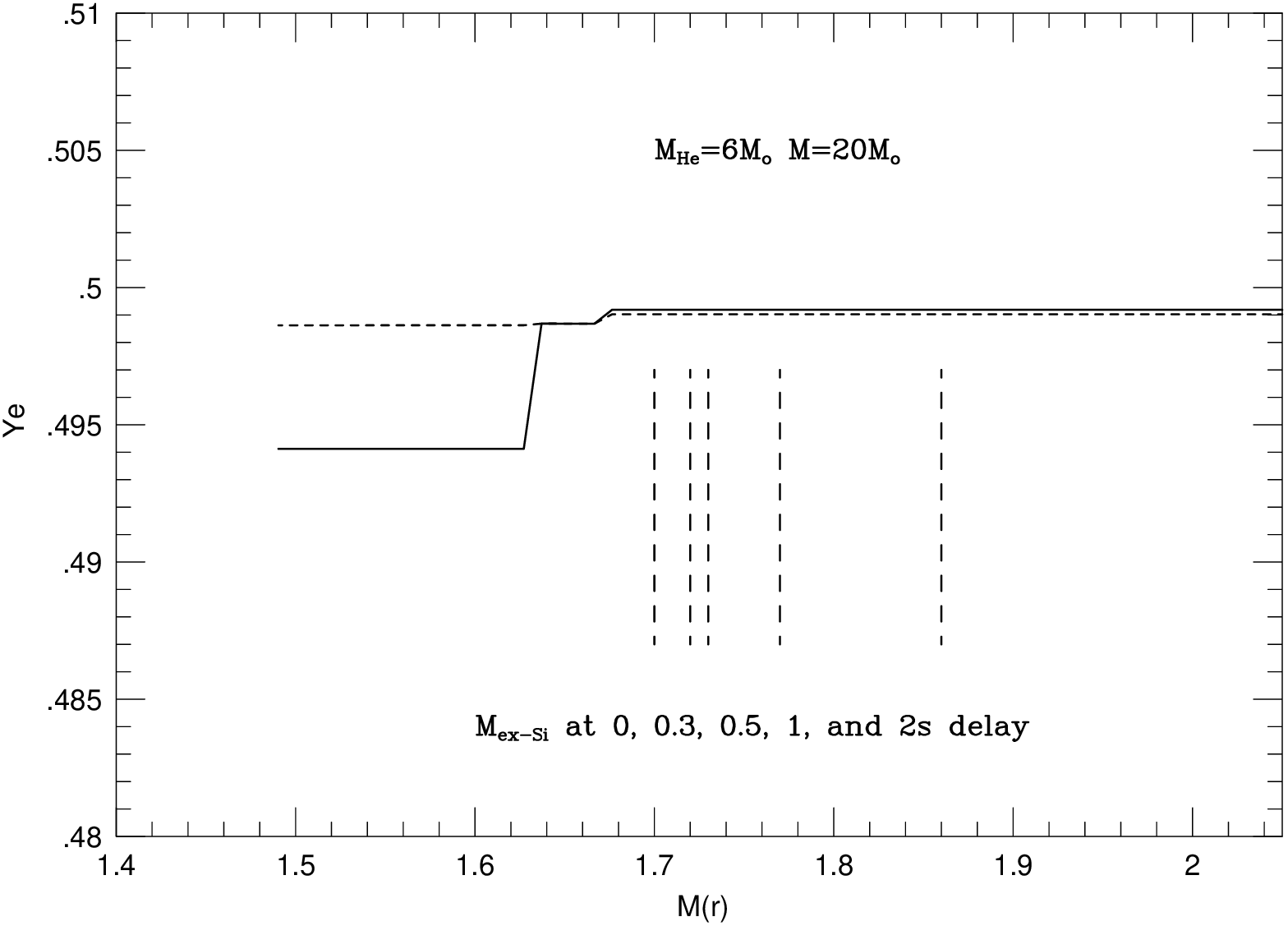,width=9cm}
\end{figure}
                                                                                
In Figs.~\ref{accneut}ab we display the $Y_e$-distributions of a 13 and a 
20M$_\odot$ star and the position of the outer boundary of explosive   
Si-burning with complete Si exhaustion, $M_{ex-Si}$, as a function of           
the delay time $t_{de}$.                                                        
We consider for each star delay times of 0, 0.3, 0.5, 1, and 2s, resulting      
in $r_{0,5}$=3700, 4042, 4412, 5410, and 7348km.                                
Inside this boundary $^{56}$Ni is produced as the dominant nucleus and the      
mass cuts would have to be positioned at $M_{cut}$=$M(r_{ex-Si})-               
M_{ej}(^{56}$Ni)=$M(r_{0,5}(0))-M_{ej}(^{56}$Ni). When Ni-ejecta of 0.15 and
0.07M$_\odot$ are used for 13 and 20M$_\odot$ stars, mass cuts $M_{cut}$ of 
1.27 and 1.61M$_\odot$ result for a vanishing delay time.           
For $t_{de,i}$=0.3, 0.5, 1, and 2s the accreted masses $\Delta M_{acc,i}$ of
0.02, 0.03, 0.07-0.08, and 0.14-0.16M$_\odot$                
have to be added to $M_{cut}$. It is recognizable that       
especially for the 13M$_\odot$ star the $Y_e$'s encountered for these           
different delay times vary strongly, and differences of the Fe-group             
composition can be expected. Assuming that the stellar models are correct,
all delay times less than 1s for the 13M$_\odot$ star are not compatible
with the chemical evolution of our galaxy, as will be discussed below.
On the other extreme, the $Y_e$ in the             
innermost ejecta of a 25M$_\odot$ star are not affected at all         
by the available choices. A more detailed discussion for the 20M$_\odot$ 
star will follow.
                                                                                
The neutron star boundary would have to be moved outward, accordingly,       
by adding $\Delta M_{acc,i}$ mentioned in the previous paragraph.    
Whether a neutron star or black hole is
formed depends on the permitted maximum
neutron star mass, which is somewhat uncertain and related to the
still limited understanding of the nuclear equation of state beyond
nuclear densities (e.g. Glendenning 1991, Weber \& Glendenning 1991, Brown
\& Bethe 1994, Prakash et al. 1997).      
A proto-neutron star with a baryonic mass                                     
$M_b$ will release a binding energy                                             
$E_{bin}$ in form of black body radiation in neutrinos                          
during its contraction to neutron star densities.                               
The gravitational mass is then given by                                         
                                                                                
\begin{equation}
M_g=M_b - E_{bin}/c^2.                                            
\end{equation}
                                                                                
For reasonable uncertainties in the equation of state,                          
Lattimer and Yahil (1989) obtained a relatively tight relation between 
gravitational mass and binding energy.                                    
Applying their expression results in a gravitational mass of the formed         
neutron star $M_g$.    
An error of roughly $\pm$15\% for the difference $M_b-M_g$ applies.             
$\Delta M_{acc}$, due to the uncertainty of the accretion period or delay       
time, and the choice of $M_{ej}(^{56}$Ni) which determines $M_{cut}$, 
dominate the error in $M_g$ of 1.16+(0-0.11)+(0.15-$M_{ej}(^{56}$Ni))
for the example of the 13M$_\odot$ star and 
1.45+(0-0.12)+(0.07-$M_{ej}(^{56}$Ni))
for the 20M$_\odot$ and possible delay periods between 0 and 2s.         
The first bracket includes uncertainties in $t_{de}$, the second one
in the actually ejected $^{56}$Ni mass.
A delay time of about 1s is expected to be an upper bound                      
for the delayed explosions. This is close to a pure neutrino diffusion          
time scale without any convective turnover.                                     
                                                                                
The results indicate a clear spread of neutron star masses.          
This spread would be preserved in real supernova events, unless a possible   
conspiracy in the combination of proto-neutron star masses, delay times,        
and explosion energetics (i.e. the explosion mechanism in general) 
leads to a smaller range in                            
neutron star masses. A certain spread is also found in                       
neutron star masses from observations (e.g. Nagase 1989, Page and               
Baron 1990, van Paradijs 1991, van Kerkwijk, van Paradijs, \& Zuiderwijk 1995,
van Paradijs \& McClintock 1995, Thorsett 1996) but it is not clear to which 
extent it is just
due to large observational errors. It is possible that the range            
predicted here already includes the uncertain upper    
mass limit of neutron stars due to the nuclear equation of state                
(Baym 1991, Weber and Glendenning 1991, Prakash et al. 1997). If it does, we 
would expect for        
these cases the                                                                 
formation of a central black hole during the delay period. Thus, no      
supernova explosion would occur and no yields be ejected. Different        
maximum stable masses between the initially hot and a cold neutron star (see 
e.g. Brown \& Bethe 1994, Prakash et al. 1997 and the discussion of kaon
condensates) could result in a supernova 
explosion {\it and} afterwards the formations of a central black hole.        
Timmes, Woosley, \& Weaver (1996) had the stamina to predict a neutron star 
initial
mass function based on ideas similar to the ones presented above. This is
probably a somewhat bold undertaking, given the fact that we do not understand
the supernova explosion mechanism fully, yet, and the $\Delta M_{acc}$ can
vary widely. But there is one aspect which is worth mentioning. Due to the
temperature dependence of the $^{12}$C($\alpha,\gamma)^{16}$O rate,
stars above a critical mass limit will leave after core He-burning less than
a critical amount of $^{12}$C ($\approx$0.1M$_\odot$), which leads to
radiative rather than convective core C-burning and finally the formation of 
large Fe-cores, which probably form black holes rather than supernovae.
This would also agree with observations based on the O/Fe ratio in early
galactic evolution, which requires an upper mass limit of about
25-50M$_\odot$ in order to avoid too high a production of oxygen
(Maeder et al. 1992, Tsujimoto et al. 1997). 
                                                                                
\subsection{Observational Constraints}                         
\noindent 
There exist a number of quantitative comparisons for SN1987A                    
(a 20M$_\odot$ star during its main sequence evolution) between                 
nucleosynthesis predictions and observations [see e.g. Table 2 in               
Danziger et al. (1990), section IVb in Thielemann et al.                        
(1990) or McCray (1993), Fransson, Houck, \& Kozma (1996), Chugai (1994)], 
which show reasonable agreement for C, O, Si, Cl, Ar, Co, and Ni 
(or Fe) between observation and theory.                                         
We want to concentrate here on a crucial aspect, the O                          
abundance.                                                                      
                                                                                
The amount of $^{16}$O is closely linked to the "effective"                     
$^{12}$C($\alpha ,\gamma$)$^{16}$O rate                                         
during core He-burning. This effective rate is determined by three              
factors: (1) the actual nuclear rate,                                           
(2) the amount of overshooting, mixing fresh He-fuel into the core              
at late phases of He-burning, when the temperatures are relatively high         
and favor alpha-captures on $^{12}$C, and                                       
(3) the stellar mass or He-core size, which determine the central temperature   
during He-burning.                                                              
                                                                                
We have discussed above
the nuclear rate and its uncertainties and used in Thielemann et al. (1996a)   
the rate by Caughlan et al. (1985) based on an astrophysical     
S-factor of $S_{tot}(0.3$MeV)=0.24MeV barn. The              
S-factor is composed of an E1 component in the range 0.08$\pm$0.020 and 
an E2-component with a much larger uncertainty of 0.066-0.064+0.104~MeV barn,
thus ranging in total from 0.062 to 0.270~Mev barn
(the references where quoted in section 3). 
As the rate by Caughlan et al. (1985) seems to 
be close to the upper limit, it is crucial to check the observations for 
individual stellar models, in order to normalize the O-production correctly.  
The model calculations for a 20 M$_\odot$ star predict 1.48M$_\odot$ of ejected 
$^{16}$O. This is within observational constraints by             
Franson, Houck \& Kozma  (1993) who found about 1.5M$_\odot$ and         
Chugai (1994) who determined 1.2-1.5M$_\odot$. It should, however,         
be clear that these observations test only the combined effect of               
nuclear rate and convection treatment (here Schwarzschild without               
overshooting). Similar results were found by Werner et al. (1995) when          
analyzing spectra of young white dwarfs with models of d'Antona and             
Mazzitelli (1992).                                                              
                                                                                
The O-determinations for SN 1993J from Houck \& Fransson (1996) result
in $\approx$0.5M$_\odot$. Thielemann et al. (1996a) predicted 
0.423M$_\odot$ for a 15M$_\odot$ main sequence star, which agrees fairly well,
SN 1993J was determined to be a 14$\pm$1M$_\odot$ star.       
This leads to the conclusion that the Caughlan et al.        
(1985) rate, used in conjunction with the Schwarzschild criterion for convection
and no overshooting, gives a very good agreement with observations for 
individual supernovae. Only a comment about this combined usage can be made.
Statements about the $^{12}$C($\alpha,\gamma)^{16}$O rate alone, in addition
based on hydrostatic rather than explosive yields (Weaver \& Woosley 1993)
should be taken with some caution, and progress should preferably
be made by improving nuclear cross sections and stellar convection 
treatment independently.
                                                                                
Recently other diagnostics became available for abundance determinations in     
supernova remnants. In that case the progenitor mass is not known, but the      
relative abundance ratios between different elements can be tested for          
consistency with abundance  predictions for a variety of progenitor masses.     
Hughes and Singh (1994) made use of X-ray spectra of the supernova remnant      
G292.0+1.8 and found remarkable agreement for all element ratios from O through 
Ar with our 25M$_\odot$ calculations (15\% rms deviation). This tests 
implicitely the effective
$^{12}$C$(\alpha,\gamma)^{16}$O-rate, as it is also reflected in the ratios 
between C-burning products like Ne and Mg and explosive O-burning products      
like Ar and S. Comparisons with other model predictions (Woosley \& Weaver 
1995) led to larger deviations.     
UV and optical observations of supernova remnant N132D by Blair, Raymond        
\& Long (1994) give very good agreement with our element predictions for a      
20M$_\odot$ star, with slight deviations for Mg. Thus, we have direct 
observations 
of supernovae and supernova remnants ranging from 15 over 20 to 25M$_\odot$,    
which agree well with our model predictions and indicate that their application 
for other purposes should be quite reliable.                                    
This has recently also been demonstrated for galactic chemical evolution
calculations (Tsujimoto et al. 1995, 1997, Timmes, Woosley \& Weaver 1995,
Pagel \& Tautvaisiene 1995,1997).
                                                                                
The formation of the nuclei $^{58,61,62}$Ni, which are produced in form         
of the neutron-rich species $^{58}$Ni and $^{61,62}$Zn,                         
is strongly dependent on $Y_e$ and varies                                       
therefore with the position of the mass                                         
cut between ejected matter and the remaining neutron star                       
(see the discussion in Thielemann et al. 1990 and Kumagai et al. 1991, 1993).   
Especially for the Ni-abundances the position of                                
the mass cut is crucial. The $^{57}$Ni/$^{56}$Ni ratio                          
is correlated with the abundances of stable Ni isotopes, predominantly          
$^{58}$Ni, i.e. with $^{58}$Ni/$^{56}$Ni.                                       
Light curve observations of SN1987A (Elias et al. 1991, Bouchet et al. 1991, 
Suntzeff et al. 1992) could be interpreted with a high 57/56 ratio of 4 times 
solar, but this would also have required too large stable Ni abundances not 
substantiated from   
observations (Witteborn et al. 1989, Wooden et al. 1993, 1997). In order to meet
the stable Ni constraints of 3-5$\times 10^{-3}$M$_\odot$                       
(Danziger et al. 1990, Witteborn et al. 1989, and Wooden et al. 1993) only an   
upper limit of 1.4-1.7 times solar is permitted                                 
for the 57/56 ratio from our results, given in detail in
Thielemann et al. (1996). This also agrees well with       
the observations by Varani et al. (1990) and $\gamma$-ray line                  
observations by GRO (Kurfess et al. 1992, Clayton et al. 1992). The apparent    
57/56 discrepancy was solved by correct light curve and spectra modeling with a
non-equilibrium treatment of the involved ionization stages                     
at late times (Fransson \& Kozma 1993). This gives a 
consistent picture           
for observations of stable Ni, light curve observations which are sensitive     
to $^{56}$Co and $^{57}$Co decay, and the $\gamma$-ray lines emitted from       
both decays. 
                                                                                
This corresponds to a $Y_e$ at the mass cut of 0.4987 within the little nitch in
Figure \ref{accneut}b. A mass cut at deeper layers, where $Y_e$        
decreases to 0.494, would imply 57/56 ratios larger than 2.5 times              
solar. A mass cut further out, implying a $Y_e$ of 0.4989 results in a          
57/56 ratio of the order of 1 times solar. This means that in order to meet     
the $Y_e$-constraint with an ejection of 0.075M$_\odot$ of $^{56}$Ni,           
we have a required delay time of 0.3-0.5s.                                      
Keeping all uncertainties of the model in mind, this can be taken as a    
support that SN 1987A did not explode via a prompt explosion,                   
and did not experience a delayed explosion with a long delay       
time $t_{de}$$>$0.5s. The latter would correspond more to a pure neutrino       
diffusion case, while this result supports the understanding that       
larger neutrino luminosities are required than in the purely diffusive
case (Herant et al. 1994; Burrows 1996; Janka \& M\"uller 1996; Mezzacappa
et al. 1997).    

$^{44}$Ti is produced as a result of a strong alpha-rich freeze-out from
explosive Si-burning, as discussed in section 3.2.4. Fig. \ref{alpharich} 
displays
nicely that $^{44}$Ti provides a measure of the entropy in the explosively
processed matter. Exactly such conditions prevail in the innermost ejecta
as can be seen in Figs.~\ref{expnuc}ab. Thus, we have another important
observational constraint besides $^{56}$Ni and $^{57}$Ni, witnessing
temperature, entropy and $Y_e$ close to the mass cut. The predictions
for $^{44}$Ti ejecta range from 2$\times 10^{-5}$ to 1.7$\times 10^{-4}$
M$_\odot$ for stars ranging from 13 to 40 M$_\odot$ (Woosley \& Weaver
1995, Thielemann, Nomoto, \& Hashimoto 1996, Nomoto et al. 1997).
Observational limits for supernova remnants have been described in 
Timmes et al.~(1996) and recent GRO, COMPTEL gamma-ray observations of
CAS A (Iyudin et al. 1994, Dupraz et al. 1997, Hartmann et al. 1997)
yield ($1.27\pm0.34\times 10^{-4}$)M$_\odot$ with the new half
life determinations between 59 and 62~y of Norman (1997), 
G\"orres et al. (1997), and Ahmad et
al. (1997). This is a nice confirmation of nucleosynthesis predictions.
Recent light curve calculations, based on the radioactive decay energies
of $^{56}$Ni, $^{57}$Ni, and $^{44}$Ti (Kozma \& Fransson 1997), when compared
with late time light curve observations of SN 1987A (Suntzeff et al.~1997), 
also come
to the conclusion of $\approx 10^{-4}$M$_\odot$ ejecta of $^{44}$Ti in good
agreement with the predictions by Thielemann et al. (1996a), who obtained
yields typically somewhat larger than Woosley \& Weaver (1995), probably
because energy deposition provides a somewhat larger entropy for the inner
layers than induced explosions with the aid of a piston.
                                                                                
Unfortunately, we do not yet have similar observational and computational       
results for other supernovae. This would be a strong test for the explosion     
mechanism as a function of progenitor mass.      
It is important to explore the whole progenitor mass range with multidimensional
explosion calculations in order to find out what $Y_e$ and entropy 
self-consistent calculations would predict         
for the inner ejecta. Taken at face value, our 13M$_\odot$ model would    
ask for a delay time $>$1s, in order to avoid pollution of the galaxy
with an unwanted Fe-group composition.

A further test for the correct behavior of the ejecta composition as a function
of progenitor mass is the comparison 
with abundances in low metallicity                                         
stars. These reflect the average SNe II composition, integrated over an initial 
mass function of progenitor stars. First individual tests were done
in Thielemann et al. (1990, 1996a). Applications to full chemical evolution 
calculations of the galaxy were performed by e.g. Tsujimoto et al. (1995), 
Timmes et al. (1995), Pagel \& Tautvaisiene (1995, 1997), and Tsujimoto et al. 
(1997) and prove to be a clear testing ground for supernova models.
A verification of 
SNe II ejecta in such a way permits a correct application in chemical evolution
calculations together with SNe Ia and planetary nebula ejecta (stars
of initial mass $M<$8M$_\odot$ which form white dwarfs and eject their
H- and He-burned envelopes).

\section{The r-Process}

The rapid neutron-capture process (r-process) leads to the production of
highly unstable nuclei near the neutron drip-line and functions
via neutron captures, $(\gamma,n)$-photodisintegrations, $\beta^-$-decays
and beta-delayed processes. Neutrino-induced reactions may also play a possible 
role. The r-process abundances witness 
the interplay between nuclear structure far from beta-stability and the 
appropriate astrophysical environment. 
Observations of heavy elements in low metallicity stars with abundances
of Fe/H being 1/1000 to 1/100 of solar give information about stellar surface 
abundances, which are the abundances of the interstellar gas from which 
stars formed early in galactic evolution. Such observations show on the one
hand an apparently completely solar r-process abundance pattern, at least
for $A$$>$130, indicating that during such early times in galactic evolution
only r-process sources and no s-process sources contributed to the production
of heavy elements (Sneden et al.~1996, Cowan et al.~1997). That is consistent 
with the picture discussed in section
3.1 of the s-process origin in low and intermediate mass stars, which set in
only at evolution times $>$$10^8$y. 

On the other hand, it is also recognized
that the r-process abundances come in with a delay of $>$$10^7$y after
Fe and O (Mathews, Bazan, \& Cowan 1992), which excludes the higher mass
SNe II as r-process sources, because such massive stars beyond 10-12M$_\odot$
have shorter evolution times.
The r-process has generally been associated with the inner ejecta of
type II supernovae [see e.g. the reviews by~\cite{cowtt91} and~\cite{meyer94}],
but also the decompression of neutron star matter was 
suggested by \cite{lattimer77,meyer89}, and \cite{eichler89} and is consistent
with the above mentioned low metallicity observations. Both these 
environments
provide or can possibly provide  high neutron densities and high temperatures.
Models trying to explain the whole r-process composition by low neutron density 
($<10^{20}\,{\rm cm^{-3}}$) and temperature ($<10^9\,{\rm K}$) environments, 
like e.g. explosive He-burning in massive stars \cite{thielemann79}, were 
clearly invalidated by \cite{blake81}.
The high entropy wind of the hot neutron star following type II supernova
explosions has been suggested as a
promising site for r-process nucleosynthesis by \cite{woosley92,woosley94b},
and \cite{takahashi94}. 

Actual r-process calculations usually followed two different approaches.
Some studies, focusing mostly on nuclear physics issues far from stability,
made use of a 
model-independent approach for the r-process as a function of neutron number
densities $n_n$ and temperatures $T$, extending for a duration time $\tau$
[see e.g. \cite{kratzea88,kratzea93,thielemann94a,chen95},
Bouquelle et al.~(1996), Pfeiffer et al. (1997), and \cite{kratzea97}]. 
Other studies usually stayed closer to a specific astrophysical environment
and followed the expansion of
matter on expansion timescales $\tau$ with an initial entropy $S$,
passing through declining temperatures and densities until the freeze-out
of all reactions [see e.g. Woosley and Hoffman (1992), 
\cite{meyer92,howard93,hoffman96,qian96b}, \cite{hoffman97}, Meyer \& Brown
(1997ab), and~\cite{surman97}].
Here we compare the similarities and differences between the two approaches
and whether there actually exists a one-to-one relation. 
Special emphasis is given to constraints, resulting from a comparison with 
solar r-process abundances in either approach, on nuclear properties far from 
stability. In addition, 
investigations are presented to test whether some features can also provide 
clear constraints on the permitted astrophysical conditions.
This relates mostly to the $A<110$ mass range, where
the high entropy scenario in supernovae faces problems.

\subsection{Model-Independent Studies}
\label{modelindependent}

The sequence of neutron captures, $(\gamma,n)$-photodisintegrations
and beta-decays (and possibly additional reactions like beta-delayed
neutron emission, fission etc.) have in principle to be followed with a
detailed reaction network, given by a system of (several thousand) coupled 
differential equations with a dimension equal to the number of isotopes. 
This can be done efficiently, as shown in~\cite{cowtt91}, however, 
approximations 
are also applicable for neutron densities and temperatures well in 
excess of $n_n > 10^{20}\,{\rm cm^{-3}}$ and  $T>10^9\,{\rm K}$,
which cause reaction timescales as short as $\approx 10^{-4}\,{\rm s}$ 
[see \cite{cameron83,bouquelle96}, and \cite{gorarn96}]. 
As the beta-decay half-lives are longer, roughly of
the order of $10^{-1}\,{\rm s}$ to a few times $10^{-3}\,{\rm s}$, an 
equilibrium can  set in for neutron captures and photodisintegrations.
Such conditions allow to make use of the "waiting point
approximation", sometimes also called the "canonical
r-process", which is equivalent to an 
$(n,\gamma)-(\gamma,n)$-equilibrium [$n_n\langle \sigma v 
\rangle_{n,\gamma}^{Z,A}Y_{(Z,A)}=
\rho N_A \langle \sigma v
\rangle_{n,\gamma}^{Z,A}Y_nY_{(Z,A)}=
\lambda_{\gamma,n}^{Z,A+1}Y_{(Z,A+1)}$, see Eq.(1.11)] for all nuclei in an 
isotopic chain with
charge number $Z$. As the photodisintegration rate 
$\lambda_{\gamma,n}^{Z,A+1}$ is related to the capture rate
$\langle \sigma v \rangle_{n,\gamma}^{Z,A}$ by detailed balance and proportional to the
capture rate times $\exp{(-Q/kT)}$, as shown in Eq.(1.7), the maximum abundance
in each isotopic chain 
(where $Y_{(Z,A)}\approx Y_{(Z,A+1)}$) is located at the same
neutron separation energy $S_n$, 
being the neutron-capture $Q$-value of nucleus $(Z,A)$.
This permits to express the location of the "r-process path", i.e. the 
contour lines of
neutron separation energies corresponding to the maximum in all isotopic chains,
in terms of the neutron number density
$n_n$ and the temperature $T$ in an astrophysical environment,
when smaller effects like ratios of 
partition functions are neglected, as reviewed in~\cite{cowtt91}.

The nuclei in such r-process paths, which are responsible for the solar 
r-process abundances, are highly neutron-rich, unstable, and located 
$15-35$ units away from $\beta$-stability with neutron separation energies of 
the
order $S_n=2-4~{\rm MeV}$. These are predominantly nuclei not accessible in 
laboratory experiments to date. The exceptions in the $A=80$ and 130 peaks
were shown in~\cite{kratzea88} and \cite{kratzea93} and continuous efforts
are underway to extend experimental information in these regions of the closed
shells $N$=50 and 82 with radioactve ion beam facilities.
The dependence on nuclear masses or mass model predictions enters via 
$S_n$. The beta-decay properties along contour lines of constant $S_n$
towards heavy nuclei [see e.g.~Fig.~4 in ~\cite{thielemann94a} or Fig.~12 below
for the region around the $N$=82 shell closure] are responsible
for the resulting abundance pattern. The build-up of heavy nuclei
is governed within the waiting point approximation only by effective
decay rates $\lambda_{\beta}^Z$ of isotopic chains.
Then the environment properties $n_n$ and $T$ (defining
the $S_n$ of the path), and the  duration time $\tau$, predict the 
abundances.
In case the duration time $\tau$ is larger than the longest half-lives
encountered in such a path, also a steady flow of beta-decays will follow,
making the abundance ratios independent of $\tau$
($\lambda_\beta^Z Y_{(Z)}=const.$ for all $Z$'s, where $Y_{(Z)}$ is the total
abundance of an isotopic chain and $\lambda_\beta^Z$ its effective
decay rate).
\begin{figure}[htb]
\epsfig{file=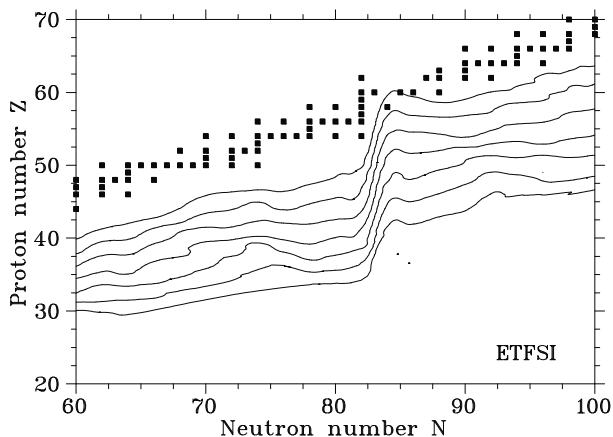,width=8cm}
\caption[F1]{Contour plots of constant neutron separation energies 
$S_n$=1,2,3,4,5,6, and 7~MeV in the
$80 \leq A \leq 140$ mass region for the ETFSI mass model~\cite{aboussir95}.
The saddle point behavior before the shell closure 
at $N=82$, also existing when using the FRDM masses by~\cite{moeller95}, causes 
a deep trough before the peak at $A=130$ (see upper part of Fig.~13), as the 
step from  the abundance
maximum of an isotopic chain $Z$ to $Z+1$ can also cause a large jump in $N$
or equivalently $A$, leading to a large number of unpopulated mass numbers 
$A$. 
\label{contourplot} }
\end{figure}

One has to recognize a number of idealizations in this picture. It assumes a
constant $S_n(n_n,T)$ over a duration time $\tau$. Then the nuclei will still 
be existent in form of highly unstable isotopes, which have to decay back to 
beta-stability. In reality $n_n$ and $T$ will be time-dependent. As long as
both are high enough to ensure the waiting point approximation, this is not
a problem, because the system will immediately adjust to the new equilibrium
and only the new $S_n(n_n,T)$ is important. The prominent question is whether 
the decrease from equilibrium conditions in $n_n$ and $T$ (neutron freeze-out),
which initially ensure 
the waiting point approximation, down to conditions where the competition of 
neutron captures and beta-decays has to be taken into account explicitely, will
 affect the abundances strongly. In our earlier investigations we considered
a sudden drop in $n_n$ and $T$, leading to a sudden "freeze-out" of this
abundance pattern, and only beta-decays and also beta-delayed properties 
[neutron emission and fission] have to be taken into account for the final 
decay back to stability [see e.g. the effect displayed in Fig.~9 of 
\cite{kratzea93}].
\begin{figure}[htb]
\epsfig{file=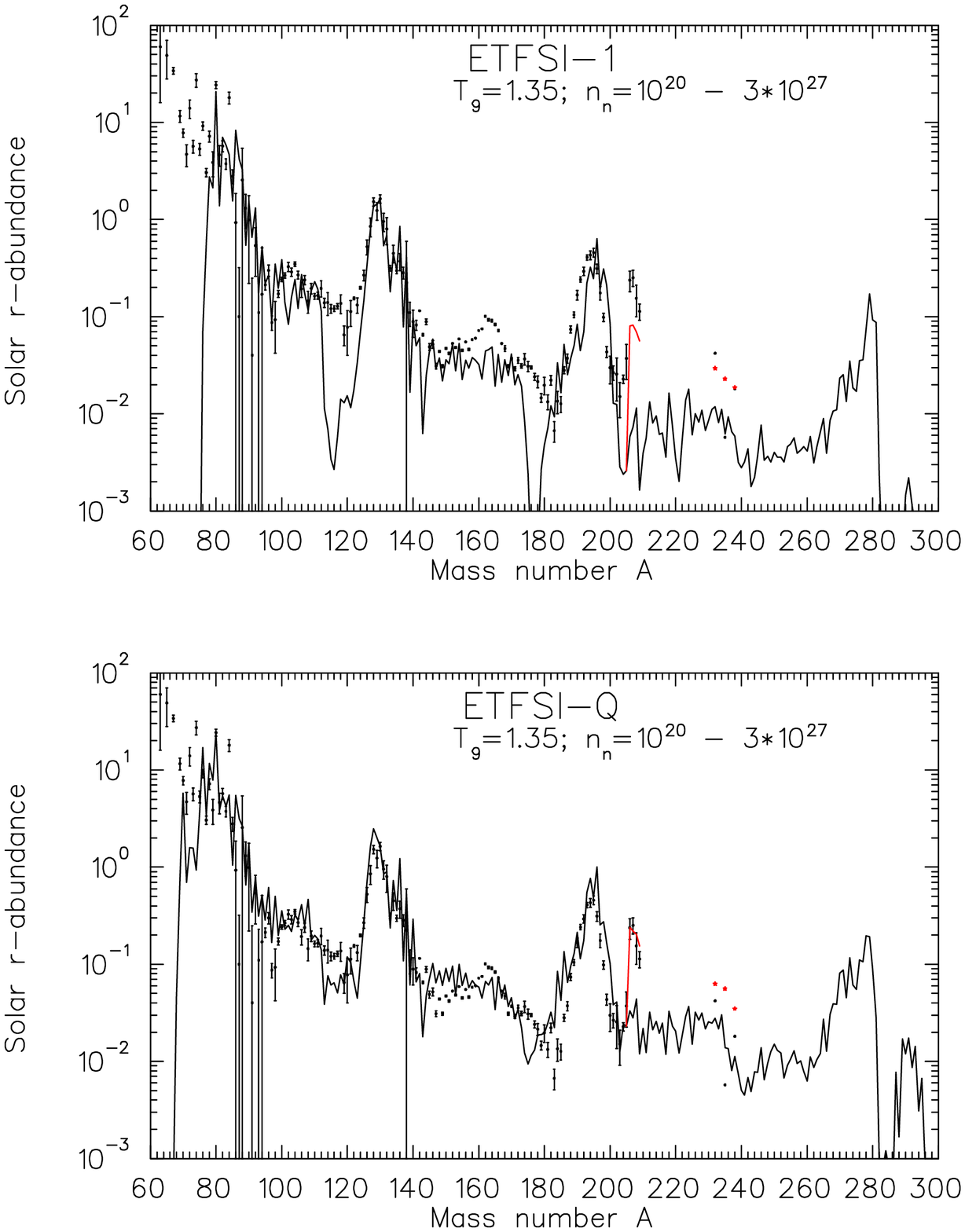,height=12cm}
\caption[F2]{Fits to solar r-process abundances by~\cite{kaeppeler89},
obtained with 17 equidistant $S_n(n_n,T)$ components from 1 to 4 MeV.
In the upper part, the result is presented for ETFSI masses with half-lives 
$\tau_{1/2}$ and beta-delayed neutron emission $P_n$ values from QRPA 
calculations.
In the lower part, the ETFSI-Q mass model by \cite{pearson96} was applied, 
which introduced a phenomenological quenching of shell effects, comparable
to HFB calculations with the Skyrme force SkP of~\cite{dobaczewski95}. 
The quenching of the $N=82$ shell gap leads to a
filling of the abundance troughs 
and to a better overall reproduction of the heavy mass region. These results
by~\cite{pfeiffer97} are also the first which show a good fit to the r-process
Pb and Bi contributions after following the decay chains of unstable heavier
nuclei. For $^{232}$Th, $^{235,238}$U the solar and r-process production
abundances are shown, allowing apparently for a increasing amount of decay
with decreasing decay half-lives (in the sequence 232, 238, 235).
\label{ueberlagerung}}
\end{figure}

When following this strategy, the analysis of the solar-system isotopic 
r-process abundance pattern showed that a minimum of three components
with different $S_n$'s, 
characterizing different r-process paths, was necessary for correctly 
reproducing the three peaks at $A\simeq80$, 130, and 195 and the 
abundances in between [\cite{thielemann93a,kratzea93}]. The "low-A wings" of the 
peaks (when making use of experimental beta-decay properties at the magic 
neutron numbers $N=50$ and 82),   
as well as the abundance pattern down to the next peak, could be 
reproduced, even with the assumption of a steady flow of beta-decays. This 
indicates that the astrophysical duration timescales $\tau$ are large in 
comparison to most of the 
beta-decay half-lives encountered and only comparable to the longest 
half-lives in the peaks (where the path comes closest to stability, see
e.g. a 2~MeV contour line in Fig.~12),
which control the leaking out
to larger $A$'s. A continuous superposition of components with varying 
$n_n$, $T$ or $S_n(n_n,T)$ (rather than only 
three), as expected in an astrophysical environment, with equidistant steps in 
$S_n$ between 2 and 4~MeV and $\tau$ between 1 and 2.5~s
led to a slight, but not dramatic, change/improvement of the abundance 
curve in \cite{kratzea94}.

When the calculations of \cite{kratzea93} were supplemented by use of the most 
modern mass formula data [Finite Range Droplet Model FRDM by \cite{moeller95} 
and Extended Thomas-Fermi model with Strutinski Integral
ETFSI by Aboussir et al.~(1995), instead of using
a somewhat dated but still very successful droplet model by
Hilf, von Groote, \& Takahashi (1976), we could show that 
abundance troughs appeared before (and after) the 130 and 195 abundances peaks,
due to the behavior of
the $S_n$ contour lines of these mass models [\cite{thielemann94a,chen95}]. 
The location in $N$ of an r-process path with a given $S_n$ does not behave 
smoothly as a function 
of $Z$. Fig.~\ref{contourplot} indicates a sudden 
jump to the position of the magic neutron number, where the contour lines
show a saddle point behavior for the FRDM as well as ETFSI mass models. The 
population gap of nuclei as a function of $A$  leads after decay to the 
abundance trough of Fig.~\ref{ueberlagerung}. 
The upper part of Fig.~\ref{ueberlagerung} shows the abundance curve 
obtained with 
ETFSI [\cite{aboussir95}] nuclear masses and beta-decay properties from a
quasi-particle random-phase approximation [QRPA, \cite{moeller97}].
When using FRDM masses by~\cite{moeller95}
instead of the ETFSI predictions, a similar picture is obtained as shown
in~\cite{thielemann94a} and \cite{bouquelle96}.

Additional tests were performed in order to see how this pattern could be
avoided with different nuclear structure properties far from stability.
The problem could be resolved in \cite{chen95}, if for very neutron-rich nuclei
the shell gap at the magic neutron number
$N=82$ is less pronounced, i.e. quenched,
than predicted by the global macroscopic-microscopic mass models.
In light nuclei, the quenching of shells in neutron-rich
isotopes is well established and a long-studied 
effect [see~\cite{orr91,campi75,fukunishi92}, and~\cite{sorlin93}].
The Hartree-Fock-Bogoliubov 
calculations by~\cite{werner94}, \cite{dobaczewski94}, and~\cite{dobaczewski95}
with a specific Skyrme force had 
exactly the expected effect on the r-process path and the resulting abundance 
curve, as shown in~\cite{chen95}.
This effect was recently also confirmed by \cite{pearson96},
when the ETFSI mass formula was phenomenologically quenched in a similar way 
as the HFB results and led to a very good
agreement with solar r-abundances in a more systematic study 
by~\cite{pfeiffer97} shown in the lower part of Fig. \ref{ueberlagerung}.
An experimental investigation of shell quenching
along the $N=50$ and 82 shell towards more neutron-rich nuclei (and approaching
the r-process path for $N=126$) is a highly desirable
goal in order to test the nuclear structure responsible for the solar
abundances of heavy nuclei. 

There are two aspects which have to be considered when trying to relate
these simplified, model-independent results to astrophysics: (a) what kind of
environments can produce the required conditions, and (b) do the 
nuclear structure conclusions drawn from the sudden freeze-out approximation 
stay valid for actual freeze-out
timescales encountered in a specific environment? The second
question cannot be answered in general, but only case by case. The 
question  whether we understand fully all astrophysical sites leading to
an r-process is not a settled one. There are strong indications that it is 
associated with type II supernovae. But galactic 
evolution timescales indicate that these can probably
only be the low mass SNe II with longer evolution timescales
\cite{cowtt91}, Mathews et al. (1992), while neutron star mergers 
or still other sites are not necessarily
excluded~\cite{lattimer77,meyer89,eichler89}.

\subsection{Parameter Studies for High Entropies}
\label{pstudy}

\subsubsection{The Model and Nuclear Input}
\label{sectionmodel}
Recent r-process studies by~\cite{woosley94b,takahashi94}, Qian 
\& Woosley (1996), and \cite{hoffman97} have concentrated on  
the hot, neutron-rich environment in the innermost ejecta of type-II 
supernovae, also called the neutrino wind. These are the layers heated by 
neutrino emission and evaporating from the
hot proto-neutron star after core collapse. These calculations obtain neutron 
separation energies of the r-process path $S_n$ of $2-4\, {\rm MeV}$, in 
agreement with the conclusions of section 
\ref{modelindependent}. Whether the entropies required for these conditions 
can really be attained
in supernova explosions has still to be verified. In relation to the questions
discussed in section \ref{modelindependent}, it also has to be investigated 
whether
a sudden freeze-out is a good approximation to these astrophysical conditions.
In order to test this, and how explosion entropies can be translated into
$n_n$ and $T$ (or $S_n$) of the model independent approach, we performed a
parameter study based on the entropy $S$ and the total proton to nucleon ratio 
$Y_e$ (which measures the neutron-richness of the initial composition), in
combination with an 
expansion timescale (for the radius of a blob of matter) of typically 0.05~s 
as in~\cite{takahashi94}, and
varied nuclear properties (i.e. mass models) like in section 
\ref{modelindependent}.

Thus, a hot blob of matter with entropy $S$, (i) initially consisting of 
neutrons, protons and some alpha-particles in NSE ratios given by $Y_e$, expands
adiabatically and cools, (ii) the nucleons and alphas combine to heavier nuclei
(typically Fe-group) with some neutrons and alphas remaining, (iii) for high
entropies an alpha-rich freeze-out from charged-particle reactions occurs for 
declining temperatures, leading to nuclei in the mass range $A\approx 80 - 100$,
and (iv) finally these remaining nuclei with 
total abundance $Y_{seed}$ can capture the remaining neutrons $Y_n$ and undergo
an r-process.
We chose a parameterized model for the expansion, essentially to introduce an
expansion timescale, which
makes these calculations independent of any specific supernova environment. But
we will have to test later whether the expansion timescale employed is relevant
to  the supernova problem. The calculations were performed for a grid of 
entropies $S$ and electron abundances
$Y_e$ ($S=3, 10, 20, 30, 
\ldots 390 k_B/{\rm baryon}$ and $Y_e=0.29, 0.31, \ldots 0.49$).
Neutron capture rates
were calculated with the new version of the statistical model code SMOKER by
\cite{rauscher97}, discussed in section 2.
The $\beta^-$-rates came from  experimental data or QRPA calculation by
\cite{moeller97}.

\begin{figure}[htb]
\epsfig{file=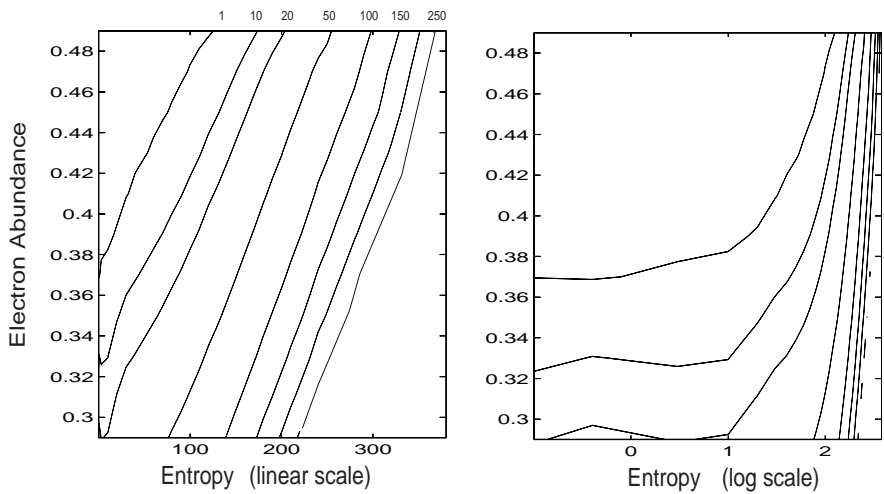,angle=0,width=13cm,
height=6.5cm}
\caption{$Y_n/Y_{seed}$ contour plots as a function of intial entropy 
$S$ and $Y_e$ for an expansion time scale of 0.05s, as expected from type II
supernova conditions.\label{ynyseedplot}}
\end{figure}

Different mass zones have different initial entropies, which leads therefore 
to a superposition of different contributions in the total ejecta. 
For each pair of parameters $Y_e$ and $S$, the calculations were initially
started with a full charged particle nuclear network up to Pd.
After the $\alpha$-rich freeze-out, an r-process network containing
only neutron induced reactions and beta-decay properties followed
the further evolution.
The dynamical r-process calculations were performed in the way as described
in Cowan, Thielemann, \& Truran (1991) and Rauscher et al. (1994).
The amount of subsequent r-processing depends on the available number 
of neutrons per heavy nucleus $Y_n/Y_{seed}$ ($Y_{seed}=\sum_{A>4}Y_{(Z,A)}$).
In Fig.~\ref{ynyseedplot}a
the $Y_n/Y_{seed}$-ratio is plotted in the $(S, Y_e)$-plane. A simple
scaling with $Y_e$ is clearly visible. Fig.~\ref{ynyseedplot}b also shows that 
low $Y_e$-values would be one mean to avoid the very high entropies required to
obtain large  $Y_n/Y_{seed}$-ratios.

\subsubsection{Superpositions of Entropies}
\label{super}
\begin{figure}[ht]
\epsfig{file=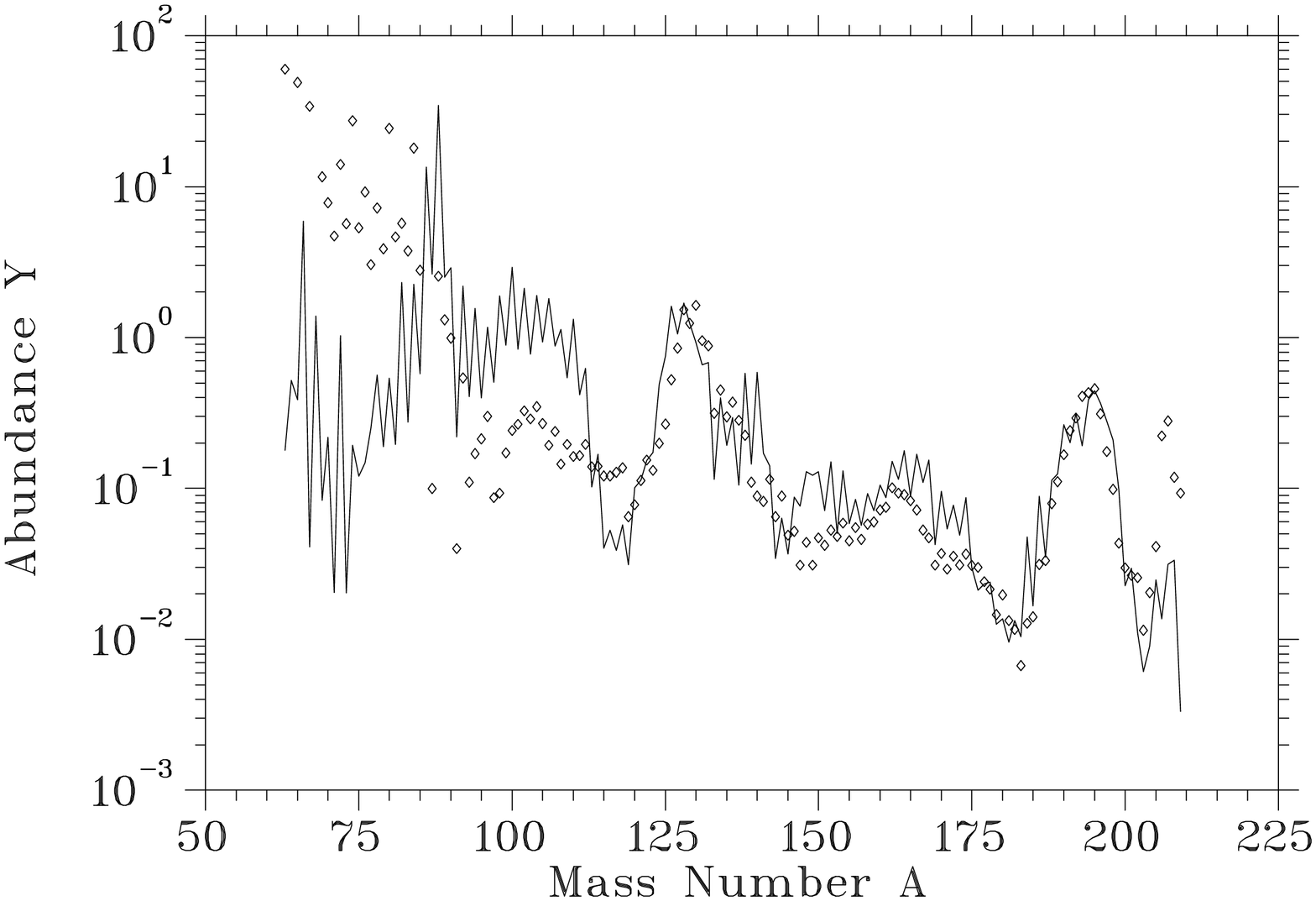,angle=90,width=8cm}
\epsfig{file=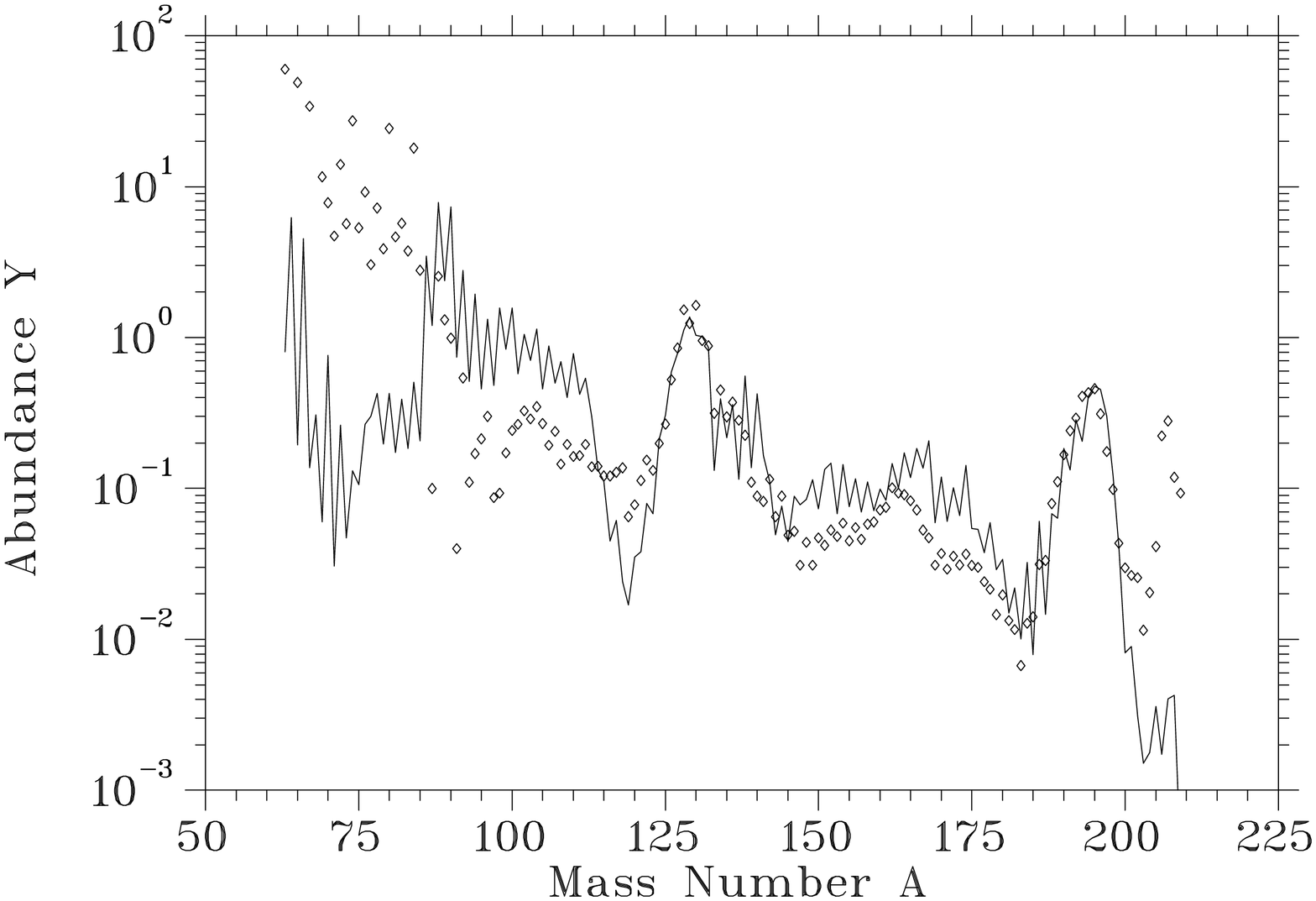,angle=90,width=8cm}
\caption[F3]
{Similar to Fig \ref{ueberlagerung} with 
the ETFSI mass formula, making use of a superposition
of entropies $g(S)$ to attain an overall good fit to solar r-process 
abundances from the high entropy neutrino wind in type II supernovae.
These calculations were performed
with $Y_e=0.45$, and 0.49, but similar results are obtained in the
range $0.30-0.49$, only requiring a scaling of entropy. The trough below
$A=130$ behaves similar to Fig.~\ref{ueberlagerung}. This shows that
a time dependent freeze-out (with a full treatment of neutron captures and
photodisintegrations) resulting from a more realistic astrophysical scenario, 
can cause the same abundance deficiencies due to specific nuclear structure 
features as obtained  in an instantaneous freeze-out from $(n,\gamma)-
(\gamma ,n)$-equilibrium. The trough before the $A=195$ peak existing e.g.
for the ETFSI mass formula in the waiting point approximation and an 
instantaneous freeze-out is filled due to non-equilibrium freeze-out
neutron captures. The strong deficiencies in the abundance pattern below 
$A=110$ are due to 
the alpha-rich freeze-out and thus related to the astrophysical scenario
rather than to nuclear structure. 
\label{ueberlagerung2}}
\end{figure}

The remaining question is, what kind of superposition of entropies the
astrophysical environment provides. The calculations of \cite{witti94} showed 
that
the amount of mass ejected per entropy interval was relatively constant at 
late phases (when the higher entropy matter was expelled) and declining
slightly at early phases (lower entropies) as a function of time (i.e. 
with increasing entropy). We did not perform complete hydro calculations, but
rather followed these findings within a parametrized way, which allows to 
optimize for the best possible fit to the solar abundance distribution
via a weighting function $g(S_i)$  with $g(S_i)=x_1e^{-x_2S_i}$,
where $i$ is the index of the components.
We restrict ourselves here to two  different $Y_e$-sequences with 
$Y_e$=0.45 and 0.49 shown in Fig. \ref{ueberlagerung2}.

Entropies from about 200 to about 350 give $Y_n/Y_{seed}$-ratios growing from 
approximately 30 to 150. The $\alpha$-rich freeze-out always produces
seed nuclei in the range $90<A<120$. This material can then be 
``r-processed'', leading to a fully neutron dominated process as discussed in
section \ref{modelindependent} and the components have a very similar 
abundance pattern in the mass range
$A=110 - 200$. Thus, it is possible for this entropy range to establish a
one-to-one correspondence for abundances obtained in r-process conditions
between entropy and expansion timescale ($S,\tau$) in one type of
calculation and a neutron separation energy of the r-path and timescale
($S_n(n_n,T),\tau$) in the calculations discussed in section
\ref{modelindependent}. The neutron separation energy $S_n$ of the r-path
is the one obtained during neutron capture freeze-out in the entropy based
calculations. This correspondence can, however, only be established for
entropies producing nuclei with $A>$110.

Matter for $A$$<$110 is a result of lower entropies with a neutron-poor and 
alpha-rich freeze-out, where the abundance of heavy nuclei is dominated
by nuclei with alpha separation energies of $\approx$6~MeV and a
$Z/A$=$Y_{e,heavy}$ of the dominating heavy nucleus after charged particle
freeze-out, resulting from 
$Y_{e,global}$=$\sum_iZ_iY_i$$\approx$$0.5X_\alpha+Y_{e,heavy}
X_{heavy}$ with mass fractions $X_i$=$A_iY_i$
[for more details see Freiburghaus et al.~(1997ab)].
None of the entropies produces an abundance peak
at charged particle freeze-out with $A<80$, leaving a sufficient amount of 
neutrons for an r-processing which would reproduce the typical neutron-induced
abundance features in the range $A=80-110$. 
A different choice of $Y_{e,global}$ (shown here for 0.45 and 0.49) can
influence that pattern somewhat in avoiding very large spikes for
$A$$\approx$90 and $N$=50 isotopes, but the overall features stay.

\begin{figure}[htb]
\epsfig{file=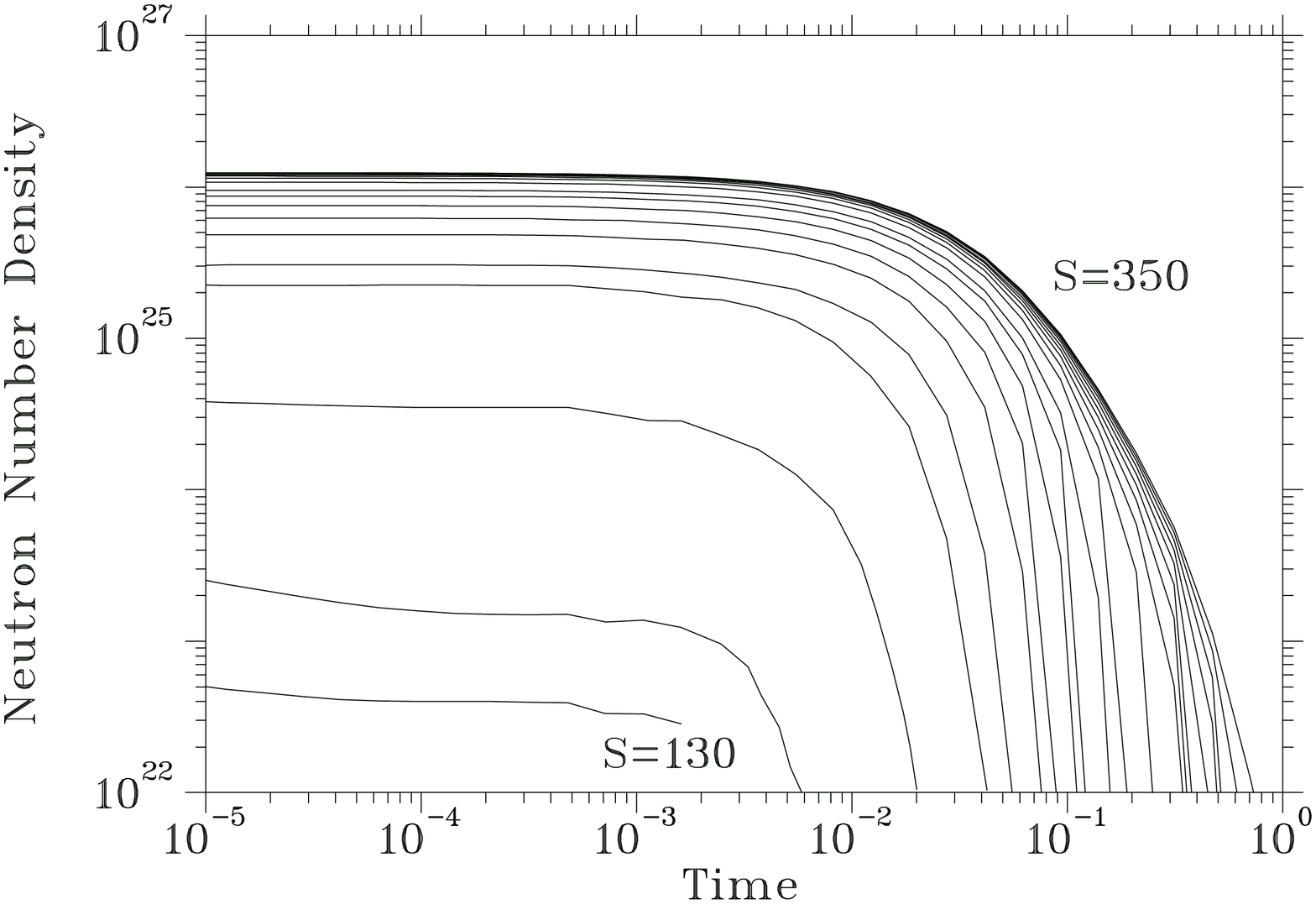,width=10cm}
\caption{$n_n(t)$ in s, displayed for different entropies $S$ in units of 
k$_B$ per baryon. \label{firstnntpicture}}
\end{figure}

Beyond $A=110$ different mass models (in Fig. \ref{ueberlagerung2} only
ETFSI is shown) give fits 
of similar quality as those displayed in section 5.1.
The discrepancies below the $A=130$ r-process
peak, in form of a pronounced trough, occur again for the FRDM and ETFSI mass
model. Thus, our results and conclusions from \ref{modelindependent} can be
translated also to "realistic" astrophysical applications for this mass
region. The nuclear
structure properties leading to agreement and deficiencies apply in the same
way, due to the nature of a fast freeze-out, which preserves the abundances
as they result from an initial $(n,\gamma)-(\gamma,n)$-equilibrium at
high temperatures, even when neutron captures and photodisintegrations are 
followed independently. Figure \ref{firstnntpicture} shows the neutron number 
densities as a function of time. Low entropies ($S=3-150$) that contribute to 
the mass range between $90<A<140$ lead to an r-process with a fast (almost 
sudden) freeze out on short timescales of $\tau \approx 0.04\,{\rm s}$. Thus it
is not surprising that the trough before $A=130$ (due to shell structure far
from stability and its effect on abundance patterns in $(n,\gamma)-(\gamma,n)
$-equilibrium) survives.

There is possibly one difference to the conclusions given with Figure 
\ref{ueberlagerung}.
As can be seen from Figure \ref{firstnntpicture}, the calculations experiencing
the highest entropies have the longest neutron freeze-out
timescales. On the other hand, they are responsible for the heaviest nuclei
with the largest neutron capture cross sections. 
Our results show that the trough before the $A=195$ peak,
resulting in case of the ETFSI mass model and a waiting point approach,
does not survive [see~\cite{thielemann94a,chen95,bouquelle96}, and 
\cite{pfeiffer97}]. This r-process 
abundance region is changed by ongoing (non-equilibrium) captures 
during the freeze-out and does not directly witness nuclear properties far
from stability at the $N=126$ shell closure. In Fig.~\ref{ueberlagerung2} 
we actually observe a filling of the minimum before the $A=195$ peak and even
the ETFSI masses, that produced the largest trough in the
waiting point calculations, seem to give a good fit.

There have been suggestions that neutrino-induced  spallation of nuclei in 
the $A=130$ peak, caused by the strong neutrino wind from the hot neutron
star, could fill the abundance trough \cite{qian96a}.
We refer to a more detailed discussion of this effect in Thielemann et al.
(1997) and Freiburghaus (1997ab), including the requirements on neutrino
luminosities and distances of matter from the neutron star at the time of
the neutron freeze-out. We come to the conclusion
that as an alternative interpretation the nuclear structure effects
(shell quenching far from stability) outlined in detail in section 
\ref{modelindependent} are still preferred, especially as they are already 
observed experimentally for lighter nuclei. 
 
What can we learn from these entropy based studies and the fact that the 
r-process abundances below $A$=110 cannot be reproduced correctly?
There are several
possible conclusions: (a) the high entropy wind is not the correct r-process
site (on the one hand due to the inherent deficiencies in the abundance 
pattern below $A=110$ and the problems to obtain the high entropies in SNe II
explosions, required for producing the massive r-process nuclei up
to $A\simeq 195$ and beyond),
or (b) the high entropy wind overcomes the problems to attain the high
entropies and produces only the masses beyond $A=110$, avoiding or
diluting the
ejection of the lower entropy matter. In the latter case another site is
responsible for the lower mass region.
An extension of $Y_e$ to smaller values, as low as 0.3, could also solve the
problem, and constraints on $\nu_e$ and $\bar{\nu}_e$ fluxes and mean energies
in the supernova environment have been explored by \cite{qian96b} to achieve
this goal. But in addition to a lower $Y_e$, also lower entropies are  required
as they might come from cold high density matter in beta-equilibrium
(see section 3.2.5 and Cameron 1989, Meyer 1989, and Hillebrandt, Takahashi,
\& Kodama 1976).
This can be deduced from Figure \ref{ynyseedplot}b. It shows the 
$Y_n/Y_{seed}$-ratio plotted in the $(S,Y_e)$-plane with a
logarithmic entropy axis and extends down to entropies as low as $S=10^{-2}$,
where a normal and not an alpha-rich freeze-out is encountered.

Whether such an interpretation ($A<130$ from low $Y_e$ and $S$ conditions, 
$A>130$ from high $S$ conditions) is the solution, might eventually be 
answered by observations. There seems to exist meteoritic evidence, discussed 
by~\cite{wasserburg96}, that the last r-process contributions
to the solar system for $A>130$ and $A<130$ came at different times, i.e.
from different types of events deduced from the extinct radioactivities
${\rm ^{107}Pd}$, ${\rm ^{129}I}$, and ${\rm ^{182}Hf}$ in meteoritic  matter.
It is highly desirable to have an independent verification of this from
observations of low metallicity stars, which apparently show a completely 
solar r-process composition for nuclei with $A>130$
[see \cite{sneden96}, and \cite{cowan97}], possibly stemming from the first 
events in our galaxy which produce r-process nuclei (Mathews et al. 1992). 
It is also necessary to explore the
abundances of nuclei with $A<130$ in such observations, in order to test 
whether the solar pattern will also be found there or is absent, due to
different evolution timescales of two independent stellar sources for these
different mass ranges of r-process nuclei.

\begin{acknowledgments}
The results presented in this study benefitted from discussions with W.D. 
Arnett, R. Azuma, W. Benz, A. Burrows, S. Bruenn, J.J. Cowan, J. Dobaczewski,
S. Goriely, 
C. Fransson, W. Hillebrandt, R. Hoffman, J. Hughes, H.-T. Janka, E. Kolbe, 
S. Kumagai, K. Langanke, J. Lattimer, A. Mezzacappa, P. M\"oller, E. M\"uller, 
M. Pearson, M. Prakash, Y.-Z. Qian, A. Ray, S. Reddy, B. Schmidt, R. Sawyer, 
F. Timmes, T. Tsujimoto, M. Turatto, P. Vogel, T. Weaver, and M. Wiescher. 
This work was supported by the Swiss Nationalfonds (grant 20-47252.96), 
the grant-in-Aid for
Scientific Research (05242102, 06233101) and COE research (07CE2002)
of the Ministry of Education, Science, and Culture in Japan,
the German BMBF (grant 06Mz864) and DFG (grant Kr80615),
the US NSF (grant PHY 9407194) and the Austrian Academy of Sciencies.
Some of us (FKT, CF, and KN) thank the ITP at the Univ. of California,
Santa Barbara, for hospitality and inspiration during the supernova program.
\end{acknowledgments}

\eject

\end{document}